\documentclass[twocolumn,aps,tightenlines,superscriptaddress,showpacs]{revtex4}
\usepackage{times}
\usepackage{amsmath}
\usepackage{graphicx}
\usepackage{epsfig}
\usepackage{threeparttable}
\usepackage{dcolumn}
\usepackage{bm}
\usepackage{multirow}
\usepackage{url}
\usepackage[colorlinks,linkcolor=red,anchorcolor=green,citecolor=blue]{hyperref}

\newcommand{\BR}{{\cal B}}

\newcommand{\beq}{\begin{equation}}
\newcommand{\eeq}{\end{equation}}
\newcommand{\bitm}{\begin{itemize}}
\newcommand{\eitm}{\end{itemize}}
\begin{document}
%\linenumbers

\preprint{} \preprint{\vbox{ \hbox{   }
%                        \hbox{Belle Preprint 2016-04}
%                        \hbox{KEK Preprint 2016-3}
                        \hbox{Intended for {\it Phys. Rev. D}}
%                        \hbox{Authors: S.~D.~Yang, C.~P.~Shen, C~.Z.~Yuan, Y.~Ban}
%                        \hbox{Committee:  Galina Pakhlova (chair), Steve Olsen and Makoto Takizawa}
}}

\title{%\quad\\[0.5cm]
\boldmath Search for $XYZ$ states in $\Upsilon(1S)$ inclusive decays}

\noaffiliation
%%%\affiliation{Aligarh Muslim University, Aligarh 202002}
\affiliation{University of the Basque Country UPV/EHU, 48080 Bilbao}
\affiliation{Beihang University, Beijing 100191}
%%%\affiliation{University of Bonn, 53115 Bonn}
\affiliation{Budker Institute of Nuclear Physics SB RAS, Novosibirsk 630090}
\affiliation{Faculty of Mathematics and Physics, Charles University, 121 16 Prague}
%%%\affiliation{Chiba University, Chiba 263-8522}
\affiliation{Chonnam National University, Kwangju 660-701}
\affiliation{University of Cincinnati, Cincinnati, Ohio 45221}
\affiliation{Deutsches Elektronen--Synchrotron, 22607 Hamburg}
%%%\affiliation{University of Florida, Gainesville, Florida 32611}
%%%\affiliation{Department of Physics, Fu Jen Catholic University, Taipei 24205}
\affiliation{Justus-Liebig-Universit\"at Gie\ss{}en, 35392 Gie\ss{}en}
%%%\affiliation{Gifu University, Gifu 501-1193}
%%%\affiliation{II. Physikalisches Institut, Georg-August-Universit\"at G\"ottingen, 37073 G\"ottingen}
\affiliation{SOKENDAI (The Graduate University for Advanced Studies), Hayama 240-0193}
\affiliation{Gyeongsang National University, Chinju 660-701}
\affiliation{Hanyang University, Seoul 133-791}
\affiliation{University of Hawaii, Honolulu, Hawaii 96822}
\affiliation{High Energy Accelerator Research Organization (KEK), Tsukuba 305-0801}
%%%\affiliation{Hiroshima Institute of Technology, Hiroshima 731-5193}
\affiliation{IKERBASQUE, Basque Foundation for Science, 48013 Bilbao}
%%%\affiliation{University of Illinois at Urbana-Champaign, Urbana, Illinois 61801}
\affiliation{Indian Institute of Science Education and Research Mohali, SAS Nagar, 140306}
\affiliation{Indian Institute of Technology Bhubaneswar, Satya Nagar 751007}
\affiliation{Indian Institute of Technology Guwahati, Assam 781039}
\affiliation{Indian Institute of Technology Madras, Chennai 600036}
\affiliation{Indiana University, Bloomington, Indiana 47408}
\affiliation{Institute of High Energy Physics, Chinese Academy of Sciences, Beijing 100049}
\affiliation{Institute of High Energy Physics, Vienna 1050}
\affiliation{Institute for High Energy Physics, Protvino 142281}
%%%\affiliation{Institute of Mathematical Sciences, Chennai 600113}
\affiliation{INFN - Sezione di Torino, 10125 Torino}
\affiliation{J. Stefan Institute, 1000 Ljubljana}
\affiliation{Kanagawa University, Yokohama 221-8686}
\affiliation{Institut f\"ur Experimentelle Kernphysik, Karlsruher Institut f\"ur Technologie, 76131 Karlsruhe}
%%%\affiliation{Kavli Institute for the Physics and Mathematics of the Universe (WPI), University of Tokyo, Kashiwa 277-8583}
%%%\affiliation{Kennesaw State University, Kennesaw, Georgia 30144}
\affiliation{King Abdulaziz City for Science and Technology, Riyadh 11442}
%%%\affiliation{Department of Physics, Faculty of Science, King Abdulaziz University, Jeddah 21589}
\affiliation{Korea Institute of Science and Technology Information, Daejeon 305-806}
\affiliation{Korea University, Seoul 136-713}
%%%\affiliation{Kyoto University, Kyoto 606-8502}
\affiliation{Kyungpook National University, Daegu 702-701}
\affiliation{\'Ecole Polytechnique F\'ed\'erale de Lausanne (EPFL), Lausanne 1015}
\affiliation{P.N. Lebedev Physical Institute of the Russian Academy of Sciences, Moscow 119991}
\affiliation{Faculty of Mathematics and Physics, University of Ljubljana, 1000 Ljubljana}
\affiliation{Ludwig Maximilians University, 80539 Munich}
%%%\affiliation{Luther College, Decorah, Iowa 52101}
\affiliation{University of Maribor, 2000 Maribor}
\affiliation{Max-Planck-Institut f\"ur Physik, 80805 M\"unchen}
\affiliation{School of Physics, University of Melbourne, Victoria 3010}
%%%\affiliation{Middle East Technical University, 06531 Ankara}
\affiliation{University of Miyazaki, Miyazaki 889-2192}
\affiliation{Moscow Physical Engineering Institute, Moscow 115409}
\affiliation{Moscow Institute of Physics and Technology, Moscow Region 141700}
\affiliation{Graduate School of Science, Nagoya University, Nagoya 464-8602}
\affiliation{Kobayashi-Maskawa Institute, Nagoya University, Nagoya 464-8602}
%%%\affiliation{Nara University of Education, Nara 630-8528}
\affiliation{Nara Women's University, Nara 630-8506}
\affiliation{National Central University, Chung-li 32054}
\affiliation{National United University, Miao Li 36003}
\affiliation{Department of Physics, National Taiwan University, Taipei 10617}
\affiliation{H. Niewodniczanski Institute of Nuclear Physics, Krakow 31-342}
%%%\affiliation{Nippon Dental University, Niigata 951-8580}
\affiliation{Niigata University, Niigata 950-2181}
%%%\affiliation{University of Nova Gorica, 5000 Nova Gorica}
\affiliation{Novosibirsk State University, Novosibirsk 630090}
%%%\affiliation{Osaka City University, Osaka 558-8585}
%%%\affiliation{Osaka University, Osaka 565-0871}
\affiliation{Pacific Northwest National Laboratory, Richland, Washington 99352}
%%%\affiliation{Panjab University, Chandigarh 160014}
\affiliation{Peking University, Beijing 100871}
\affiliation{University of Pittsburgh, Pittsburgh, Pennsylvania 15260}
%%%\affiliation{Punjab Agricultural University, Ludhiana 141004}
%%%\affiliation{Research Center for Electron Photon Science, Tohoku University, Sendai 980-8578}
%%%\affiliation{Research Center for Nuclear Physics, Osaka University, Osaka 567-0047}
%%%\affiliation{RIKEN BNL Research Center, Upton, New York 11973}
%%%\affiliation{Saga University, Saga 840-8502}
\affiliation{University of Science and Technology of China, Hefei 230026}
\affiliation{Seoul National University, Seoul 151-742}
%%%\affiliation{Shinshu University, Nagano 390-8621}
\affiliation{Showa Pharmaceutical University, Tokyo 194-8543}
\affiliation{Soongsil University, Seoul 156-743}
%%%\affiliation{University of South Carolina, Columbia, South Carolina 29208}
\affiliation{Sungkyunkwan University, Suwon 440-746}
\affiliation{School of Physics, University of Sydney, New South Wales 2006}
\affiliation{Department of Physics, Faculty of Science, University of Tabuk, Tabuk 71451}
\affiliation{Tata Institute of Fundamental Research, Mumbai 400005}
\affiliation{Excellence Cluster Universe, Technische Universit\"at M\"unchen, 85748 Garching}
%%%\affiliation{Department of Physics, Technische Universit\"at M\"unchen, 85748 Garching}
\affiliation{Toho University, Funabashi 274-8510}
%%%\affiliation{Tohoku Gakuin University, Tagajo 985-8537}
\affiliation{Department of Physics, Tohoku University, Sendai 980-8578}
\affiliation{Earthquake Research Institute, University of Tokyo, Tokyo 113-0032}
\affiliation{Department of Physics, University of Tokyo, Tokyo 113-0033}
\affiliation{Tokyo Institute of Technology, Tokyo 152-8550}
\affiliation{Tokyo Metropolitan University, Tokyo 192-0397}
%%%\affiliation{Tokyo University of Agriculture and Technology, Tokyo 184-8588}
%%%\affiliation{University of Torino, 10124 Torino}
%%%\affiliation{Toyama National College of Maritime Technology, Toyama 933-0293}
%%%\affiliation{Utkal University, Bhubaneswar 751004}
\affiliation{Virginia Polytechnic Institute and State University, Blacksburg, Virginia 24061}
\affiliation{Wayne State University, Detroit, Michigan 48202}
\affiliation{Yamagata University, Yamagata 990-8560}
\affiliation{Yonsei University, Seoul 120-749}
  \author{C.~P.~Shen}\affiliation{Beihang University, Beijing 100191} % Beihang
  \author{C.~Z.~Yuan}\affiliation{Institute of High Energy Physics, Chinese Academy of Sciences, Beijing 100049} % IHEP
  \author{Y.~Ban}\affiliation{Peking University, Beijing 100871} % Peking
% \author{A.~Abdesselam}\affiliation{Department of Physics, Faculty of Science, University of Tabuk, Tabuk 71451} % Tabuk
% \author{I.~Adachi}\affiliation{High Energy Accelerator Research Organization (KEK), Tsukuba 305-0801}\affiliation{SOKENDAI (The Graduate University for Advanced Studies), Hayama 240-0193} % KEK
% \author{K.~Adamczyk}\affiliation{H. Niewodniczanski Institute of Nuclear Physics, Krakow 31-342} % Krakow
  \author{H.~Aihara}\affiliation{Department of Physics, University of Tokyo, Tokyo 113-0033} % Tokyo
% \author{S.~Al~Said}\affiliation{Department of Physics, Faculty of Science, University of Tabuk, Tabuk 71451}\affiliation{Department of Physics, Faculty of Science, King Abdulaziz University, Jeddah 21589} % Tabuk
% \author{K.~Arinstein}\affiliation{Budker Institute of Nuclear Physics SB RAS, Novosibirsk 630090}\affiliation{Novosibirsk State University, Novosibirsk 630090} % BINP
% \author{Y.~Arita}\affiliation{Graduate School of Science, Nagoya University, Nagoya 464-8602} % Nagoya
  \author{D.~M.~Asner}\affiliation{Pacific Northwest National Laboratory, Richland, Washington 99352} % PNNL
% \author{T.~Aso}\affiliation{Toyama National College of Maritime Technology, Toyama 933-0293} % Toyama
% \author{H.~Atmacan}\affiliation{Middle East Technical University, 06531 Ankara} % METU
% \author{V.~Aulchenko}\affiliation{Budker Institute of Nuclear Physics SB RAS, Novosibirsk 630090}\affiliation{Novosibirsk State University, Novosibirsk 630090} % BINP
% \author{T.~Aushev}\affiliation{Moscow Institute of Physics and Technology, Moscow Region 141700} % MIPT
% \author{R.~Ayad}\affiliation{Department of Physics, Faculty of Science, University of Tabuk, Tabuk 71451} % Tabuk
% \author{T.~Aziz}\affiliation{Tata Institute of Fundamental Research, Mumbai 400005} % Tata
% \author{V.~Babu}\affiliation{Tata Institute of Fundamental Research, Mumbai 400005} % Tata
  \author{I.~Badhrees}\affiliation{Department of Physics, Faculty of Science, University of Tabuk, Tabuk 71451}\affiliation{King Abdulaziz City for Science and Technology, Riyadh 11442} % Tabuk
% \author{S.~Bahinipati}\affiliation{Indian Institute of Technology Bhubaneswar, Satya Nagar 751007} % IITB
  \author{A.~M.~Bakich}\affiliation{School of Physics, University of Sydney, New South Wales 2006} % Sydney
% \author{A.~Bala}\affiliation{Panjab University, Chandigarh 160014} % Panjab
% \author{V.~Bansal}\affiliation{Pacific Northwest National Laboratory, Richland, Washington 99352} % PNNL
  \author{E.~Barberio}\affiliation{School of Physics, University of Melbourne, Victoria 3010} % Melbourne
% \author{M.~Barrett}\affiliation{University of Hawaii, Honolulu, Hawaii 96822} % Hawaii
% \author{W.~Bartel}\affiliation{Deutsches Elektronen--Synchrotron, 22607 Hamburg} % DESY
% \author{A.~Bay}\affiliation{\'Ecole Polytechnique F\'ed\'erale de Lausanne (EPFL), Lausanne 1015} % Lausanne
% \author{I.~Bedny}\affiliation{Budker Institute of Nuclear Physics SB RAS, Novosibirsk 630090}\affiliation{Novosibirsk State University, Novosibirsk 630090} % BINP
  \author{P.~Behera}\affiliation{Indian Institute of Technology Madras, Chennai 600036} % IITM
% \author{M.~Belhorn}\affiliation{University of Cincinnati, Cincinnati, Ohio 45221} % Cincinnati
% \author{K.~Belous}\affiliation{Institute for High Energy Physics, Protvino 142281} % Protvino
% \author{D.~Besson}\affiliation{Moscow Physical Engineering Institute, Moscow 115409} % MEPhI
  \author{V.~Bhardwaj}\affiliation{Indian Institute of Science Education and Research Mohali, SAS Nagar, 140306} % IISERM
% ...
% Ljubljana
% \author{N.~Zwahlen}\affiliation{\'Ecole Polytechnique F\'ed\'erale de Lausanne (EPFL), Lausanne 1015} % Lausanne
% \author{O.~Zyukova}\affiliation{Budker Institute of Nuclear Physics SB RAS, Novosibirsk 630090}\affiliation{Novosibirsk State University, Novosibirsk 630090} % BINP
\collaboration{The Belle Collaboration}

\begin{abstract}

The branching fractions of the $\Upsilon(1S)$ inclusive decays
into final states with a $J/\psi$ or a $\psi(2S)$ are measured
with improved precision to be $\BR(\Upsilon(1S)\to J/\psi +
{\rm anything})=(5.25\pm 0.13(\mathrm{stat.})\pm 0.25(\mathrm{syst.}))\times
10^{-4}$ and $\BR(\Upsilon(1S)\to \psi(2S) + {\rm anything})=(1.23\pm
0.17(\mathrm{stat.})\pm 0.11(\mathrm{syst.}))\times 10^{-4}$. 
The first search for $\Upsilon(1S)$ decays into $XYZ$ states that decay
into a $J/\psi$ or a $\psi(2S)$ plus one or two charged tracks
yields no significant signals for $XYZ$ states in any of the
examined decay modes, and upper limits on their production rates
in $\Upsilon(1S)$ inclusive decays are determined.

\end{abstract}


\maketitle

%%%%%%%%%%%%%%%%%%%%%%%%%%%%%%%%%%%%%%%%%%%%%%%%%%%%%%%%%%%%%%%%%%%%%%%%%%%%%%%%%%%%%%%%%%%%%%%%%%%
%%%%%%%%%%%%%%%%%%%%%%%%%%%%%%%%%%%%%%%%%% Introduction %%%%%%%%%%%%%%%%%%%%%%%%%%%%%%%%%%%%%%%%%%%
%%%%%%%%%%%%%%%%%%%%%%%%%%%%%%%%%%%%%%%%%%%%%%%%%%%%%%%%%%%%%%%%%%%%%%%%%%%%%%%%%%%%%%%%%%%%%%%%%%%

During the past twelve years many charmoniumlike states, the
so-called ``$XYZ$" particles, have been
reported~\cite{EPJC.71.1534}. Most cannot be described
well by quarkonium potential
models~\cite{ARNRS.58.51,EPJC.71.1534,EPJC.74.2981}. Their unusual
properties have stimulated considerable theoretical interest and
various interpretations have been proposed, including tetraquarks,
molecules, hybrids, or hadrocharmonia~\cite{EPJC.71.1534,EPJC.74.2981,IJMPA.29.1430046}. 
To distinguish among these explanations, more experimental information is
needed, such as additional production processes and/or more decay
modes for these states. States with $J^{PC}=1^{--}$ can be studied
with initial state radiation in Belle's and BaBar's large
$\Upsilon(4S)$ data samples or via direct production in $e^+e^-$
collisions at BESIII. There is very little available
information on $XYZ$ production in the decays of narrow $\Upsilon$
states apart from the searches for charge-parity-even
charmoniumlike states in $\Upsilon(1S)$~\cite{PhysRevD.82.051504}
and $\Upsilon(2S)$~\cite{PhysRevD.87.071107} radiative decays.
A common feature of these $XYZ$
states is that they decay into a charmonium state such as $J/\psi$
or $\psi(2S)$ and light hadrons.
Inclusive decays of $\Upsilon(1S)$ into  $J/\psi$ and $\psi(2S)$
are observed with large branching
fractions of $(6.5\pm 0.7)\times 10^{-4}$~\cite{PhysRevD.70.072001, PhysLettB224.445} and $(2.7\pm 0.9)\times
10^{-4}$~\cite{PhysRevD.70.072001}, respectively, in which some of
the $XYZ$ states might have been produced before decaying  into $J/\psi$ or
$\psi(2S)$.

In this paper, we report a search for some of the $XYZ$ states in
$\Upsilon(1S)$ inclusive decays using the world's largest data
sample of $\Upsilon(1S)$. In these searches,
fourteen decay modes are considered:
$X(3872)$~\cite{PhysRevLett.91.262001} and
$Y(4260)$~\cite{PhysRevLett.95.142001} to $\pi^+\pi^-J/\psi$;
$Y(4260)$~\cite{PhysRevD.91.112007},
$Y(4360)$~\cite{PhysRevLett.98.212001} and
$Y(4660)$~\cite{PhysRevLett.99.142002} 
to~$\pi^+\pi^-\psi(2S)$; $Y(4260)$~\cite{Coan:2006rv} to
$K^+K^-J/\psi$; $Y(4140)$~\cite{Aaltonen:2009tz} and
$X(4350)$~\cite{PhysRevLett.104.112004} to $\phi J/\psi$;
$Z_c(3900)^\pm$~\cite{PhysRevLett.110.252001,PhysRevLett.110.252002},
$Z_c(4200)^\pm$~\cite{PhysRevD.90.112009} and
$Z_c(4430)^\pm$~\cite{PhysRevD.90.112009} to $\pi^\pm J/\psi$;
$Z_c(4050)^\pm$~\cite{PhysRevD.91.112007} and
$Z_c(4430)^\pm$~\cite{PhysRevLett.100.142001} to
$\pi^\pm\psi(2S)$; and a predicted $Z_{cs}^\pm$ state with mass
$(3.97\pm0.08)~\mathrm{GeV}/c^2$ and width
$(24.9\pm12.6)~\mathrm{MeV}$~\cite{JKPS55.424,PhysRevD.88.096014}
to $K^\pm J/\psi$.

%%%%%%%%%%%%%%%%%%%%%%%%%%%%%%%%%%%%%%%%%%%%%%%%%%%%%%%%%%%%%%%%%%%%%%%%%%%%%%%%%%%%%%%%%%%%%%%%%%%
%%%%%%%%%%%%%%%%%%%%%%%%%%%%%%% Data Sample and Belle Detector %%%%%%%%%%%%%%%%%%%%%%%%%%%%%%%%%%%%
%%%%%%%%%%%%%%%%%%%%%%%%%%%%%%%%%%%%%%%%%%%%%%%%%%%%%%%%%%%%%%%%%%%%%%%%%%%%%%%%%%%%%%%%%%%%%%%%%%%

The analysis utilizes a $5.74$~fb$^{-1}$ data sample collected at
the peak of the $\Upsilon(1S)$ resonance, containing $102\times
10^6$ $\Upsilon(1S)$ decays, and a $89.45$~fb$^{-1}$ data sample
collected off-resonance at $\sqrt{s}=10.52~\mathrm{GeV}$ that is
used to determine the levels of possible irreducible continuum
contributions. The data were collected with the Belle
detector~\cite{Abashian2002117,PTEP201204D001} operated at the
KEKB asymmetric-energy
$e^{+}e^{-}$~collider~\cite{Kurokawa20031,PTEP201303A001}. Large
Monte Carlo (MC) event samples of each of the investigated $XYZ$
modes are generated with EVTGEN~\cite{Lange2001152} to determine
signal line-shapes and efficiencies.
Both $XYZ$ meson production in $\Upsilon(1S)$
inclusive decays and their decays into exclusive final states
containing a $J/\psi(\psi(2S))$ and light hadrons are generated
uniformly in phase space.
Inclusive $J/\psi(\psi(2S))$ production is generated in the same models and 
subsequently decay  according to their known branching
fractions~\cite{ChinPhysC38.090001}; unknown decay modes are
generated using the Lund fragmentation model in
PYTHIA~\cite{JHEP2006.026}.
%with the final states that consist of $q\bar{q}$ ($q=u,~d,~s$) or $D\bar{D}$.
%An inclusive $\Upsilon(1S)$ MC event sample with an equivalent luminosity that is
%four times that of the data and produced with PYTHIA \cite{JHEP2006.026}
%is used to identify possible peaking backgrounds from $\Upsilon(1S)$ decays.

The Belle detector is a large solid angle magnetic spectrometer
that consists of a silicon vertex detector, a 50-layer
central drift chamber (CDC), an array of aerogel threshold
Cherenkov counters (ACC), a barrel-like arrangement of
time-of-flight scintillation counters (TOF), and an
electromagnetic calorimeter comprised of CsI(Tl) crystals (ECL) located
inside a superconducting solenoid coil that provides a $1.5~\hbox{T}$
magnetic field. An iron flux-return yoke located outside the coil is
instrumented to detect $K^{0}_{L}$ mesons and to identify muons.
A detailed description of the Belle detector
can be found in Refs.~\cite{Abashian2002117, PTEP201204D001}.

%%%%%%%%%%%%%%%%%%%%%%%%%%%%%%%%%%%%%%%%%%%%%%%%%%%%%%%%%%%%%%%%%%%%%%%%%%%%%%%%%%%%%%%%%%%%%%%%%%%
%%%%%%%%%%%%%%%%%%%%%%%%%%%%%%%%%%%% Event Selections %%%%%%%%%%%%%%%%%%%%%%%%%%%%%%%%%%%%%%%%%%%%%
%%%%%%%%%%%%%%%%%%%%%%%%%%%%%%%%%%%%%%%%%%%%%%%%%%%%%%%%%%%%%%%%%%%%%%%%%%%%%%%%%%%%%%%%%%%%%%%%%%%

Charged tracks from the primary vertex with $dr<2~\mathrm{cm}$ and
$|dz|<4~\mathrm{cm}$ are selected, where $dr$ and $dz$ are the
impact parameters perpendicular to and along the beam direction, respectively,
with respect to the interaction point. In addition,
the transverse momentum of every charged track in the laboratory
frame is restricted to be larger than $0.1~\mathrm{GeV/{\mathit
c}}$. Backgrounds from QED processes are significantly suppressed
by the requirement that the charged multiplicity ($N_{\rm ch}$) in
each event satisfies $N_{\rm ch}>4$~\cite{PhysRevD.70.071102}.
For charged tracks, information from different detector
subsystems including specific ionization in the
CDC, time measurements in the TOF and the response of the ACC
is combined to form the likelihood ${\mathcal L}_i$ for particle species $i$, where
$i=\pi$,~$K$ or $p$~\cite{like}. Charged tracks with
$R_{K}=\mathcal{L}_{K}/(\mathcal{L}_K+\mathcal{L}_\pi)>0.6$
are treated as kaons, while those with $R_{K}<0.4$ are considered to
be pions. With these conditions, the kaon (pion) identification
efficiency is $94\%$ ($97\%$) and the pion (kaon)
misidentification rate is about $4\%$ ($9\%$).
Candidate lepton tracks from $J/\psi(\psi(2S))$ are required to have a muon
likelihood ratio
$R_{\mu}=\mathcal{L}_{\mu}/(\mathcal{L}_{\mu}+\mathcal{L}_{K}+\mathcal{L}_{\pi})>0.1$~\cite{Abashian49169}
or an electron likelihood ratio
$R_e=\mathcal{L}_e/(\mathcal{L}_e+\mathcal{L}_{{\rm non}-e})>0.01$~\cite{Hanagaki485490}.
Furthermore, we require that a charged pion not be
identified as a muon or an electron with $R_{\mu}<0.95$ and $R_e<0.95$.

To reduce the effect of bremsstrahlung and final-state radiation,
photons detected in the ECL within a $50~\mathrm{mrad}$ cone of
the original electron or positron direction are included in the
calculation of the $e^+/e^-$ four-momentum. The
lepton-identification efficiencies for $e^\pm$ and $\mu^\pm$ are
about $98\%$ and $96\%$, respectively.

Since a final-state $J/\psi$ or $\psi(2S)$ is common to all of the
studies reported here, we first select either a $J/\psi$ via its
$\ell^+\ell^-$ ($\ell=e$ or $\mu$) decay mode or a $\psi(2S)$
decaying into $\ell^+\ell^-$ or $\pi^+ \pi^- J/\psi$.
For $\psi(2S)\to \pi^+ \pi^- J/\psi$, a mass-constrained fit is applied
to the $J/\psi$ candidate.

\begin{figure*}[htbp]
    \includegraphics[width=0.3\textwidth]{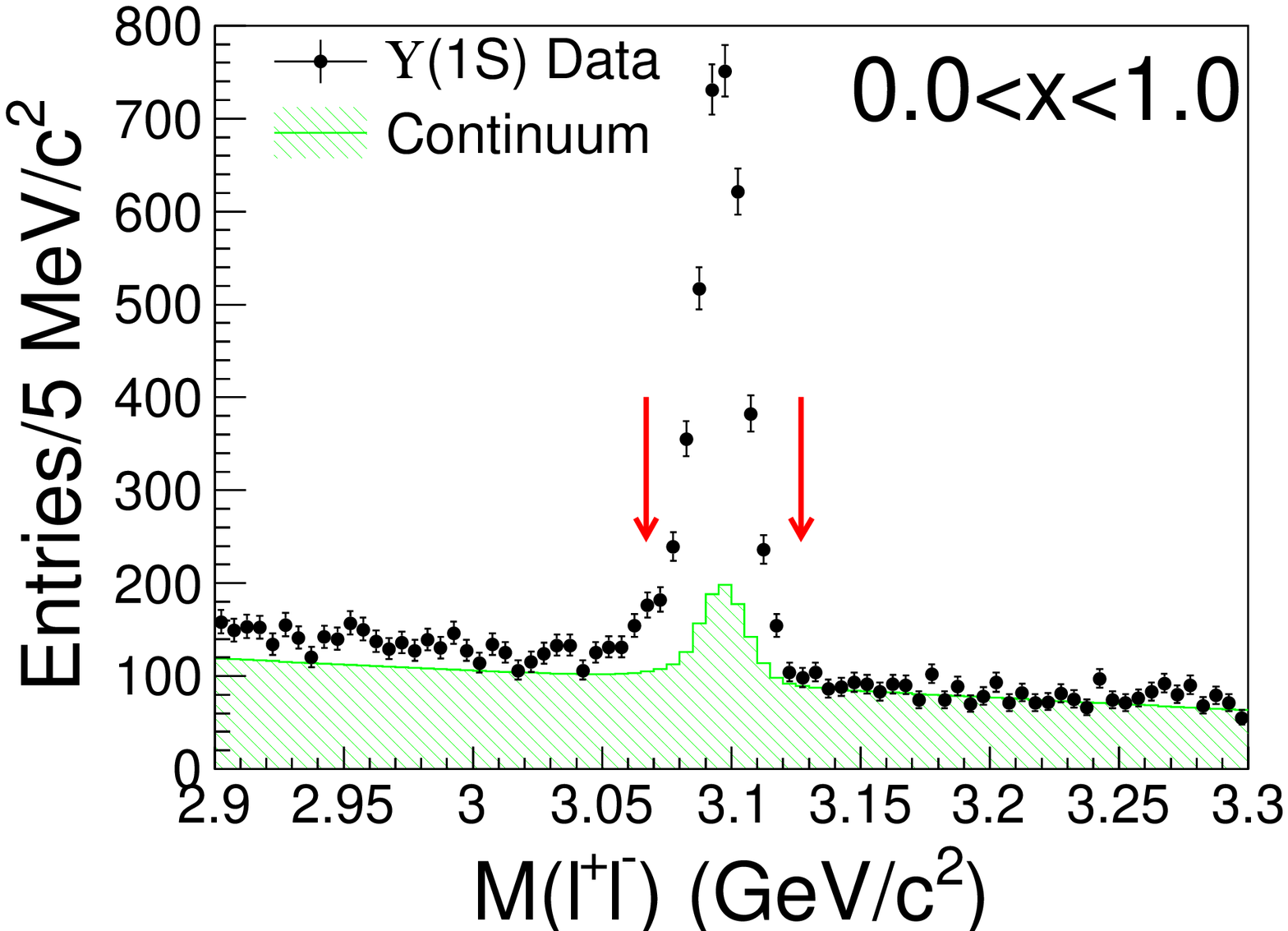}
    \includegraphics[width=0.3\textwidth]{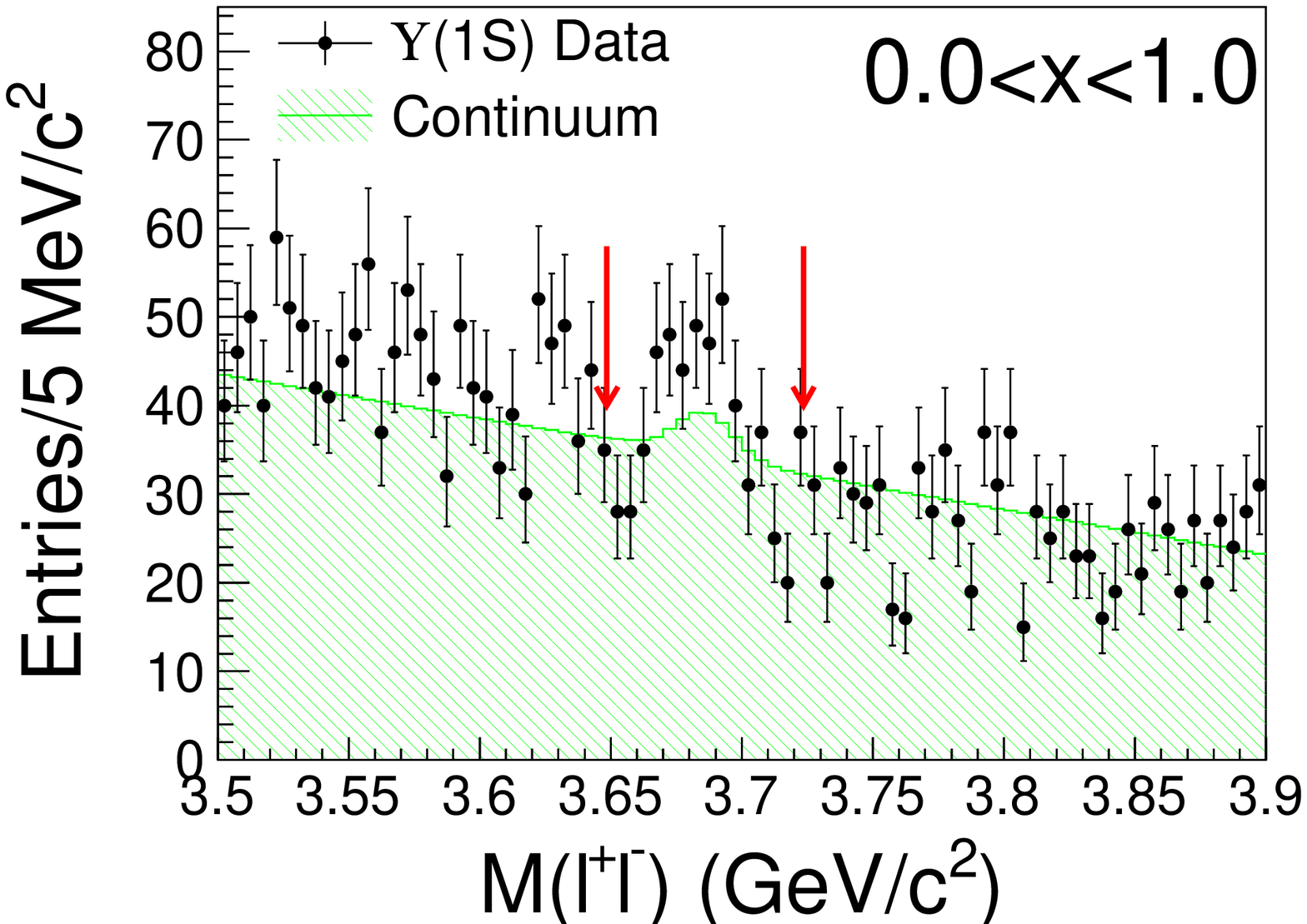}
    \includegraphics[width=0.3\textwidth]{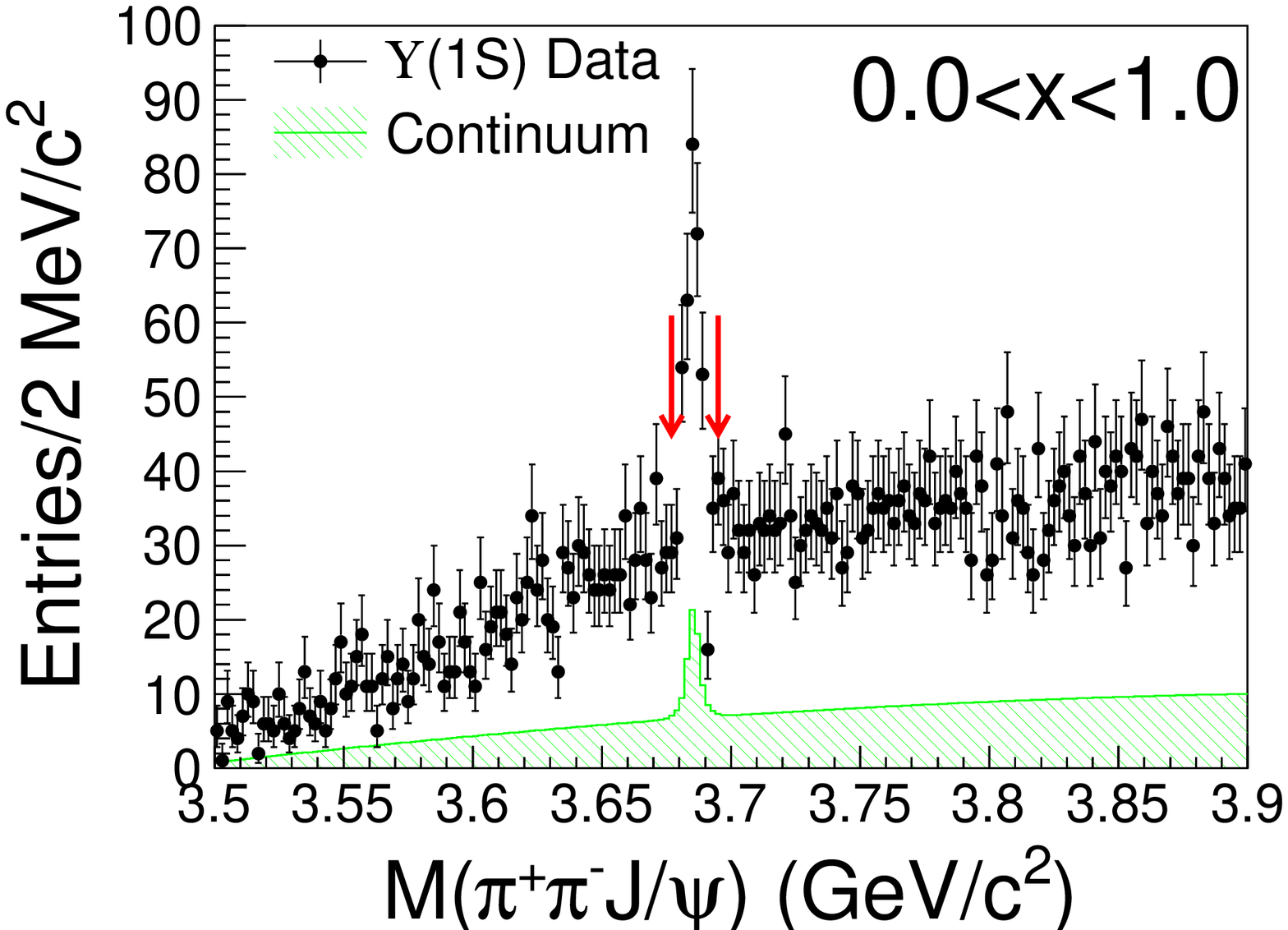}
    \includegraphics[width=0.3\textwidth]{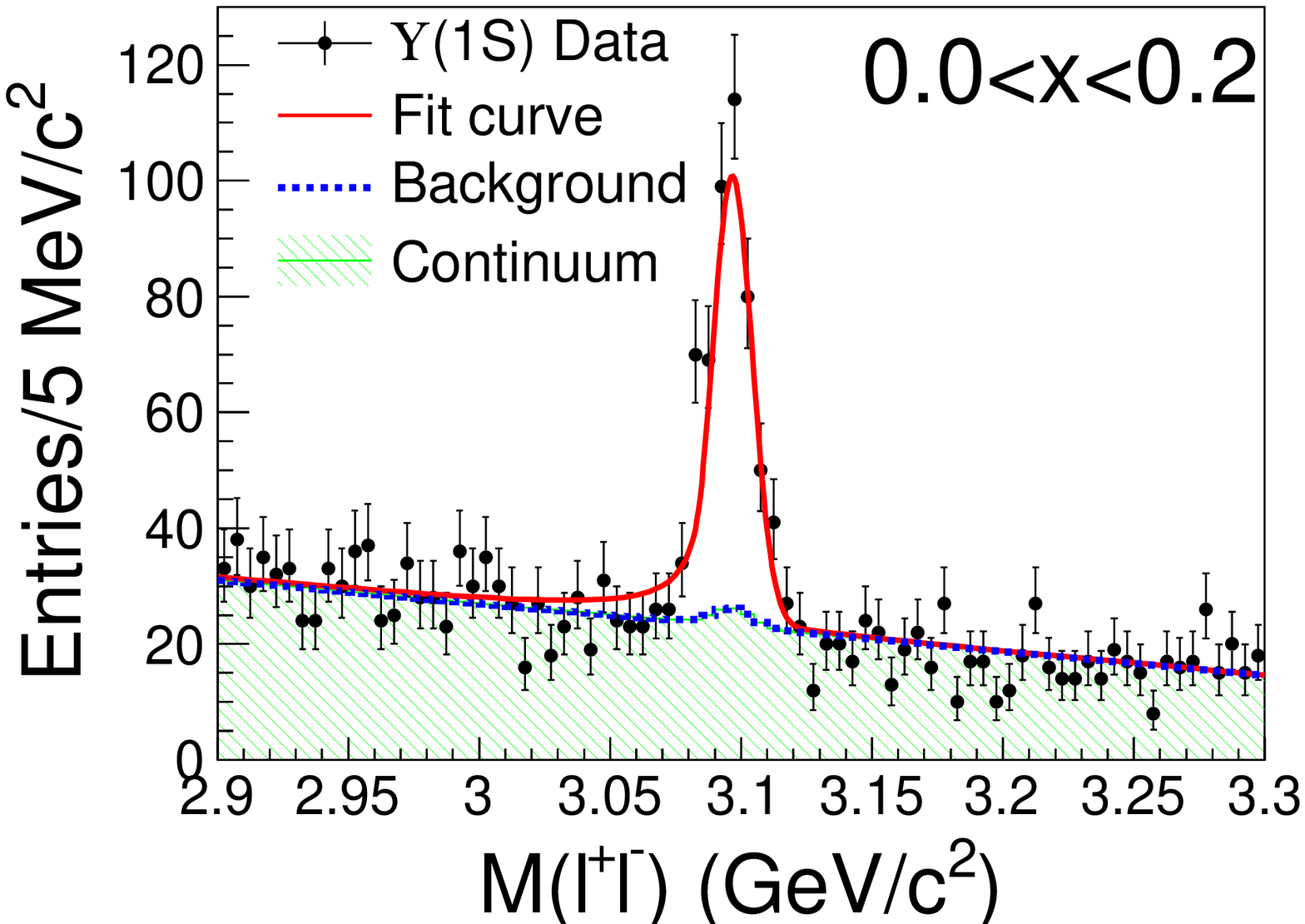}
    \includegraphics[width=0.3\textwidth]{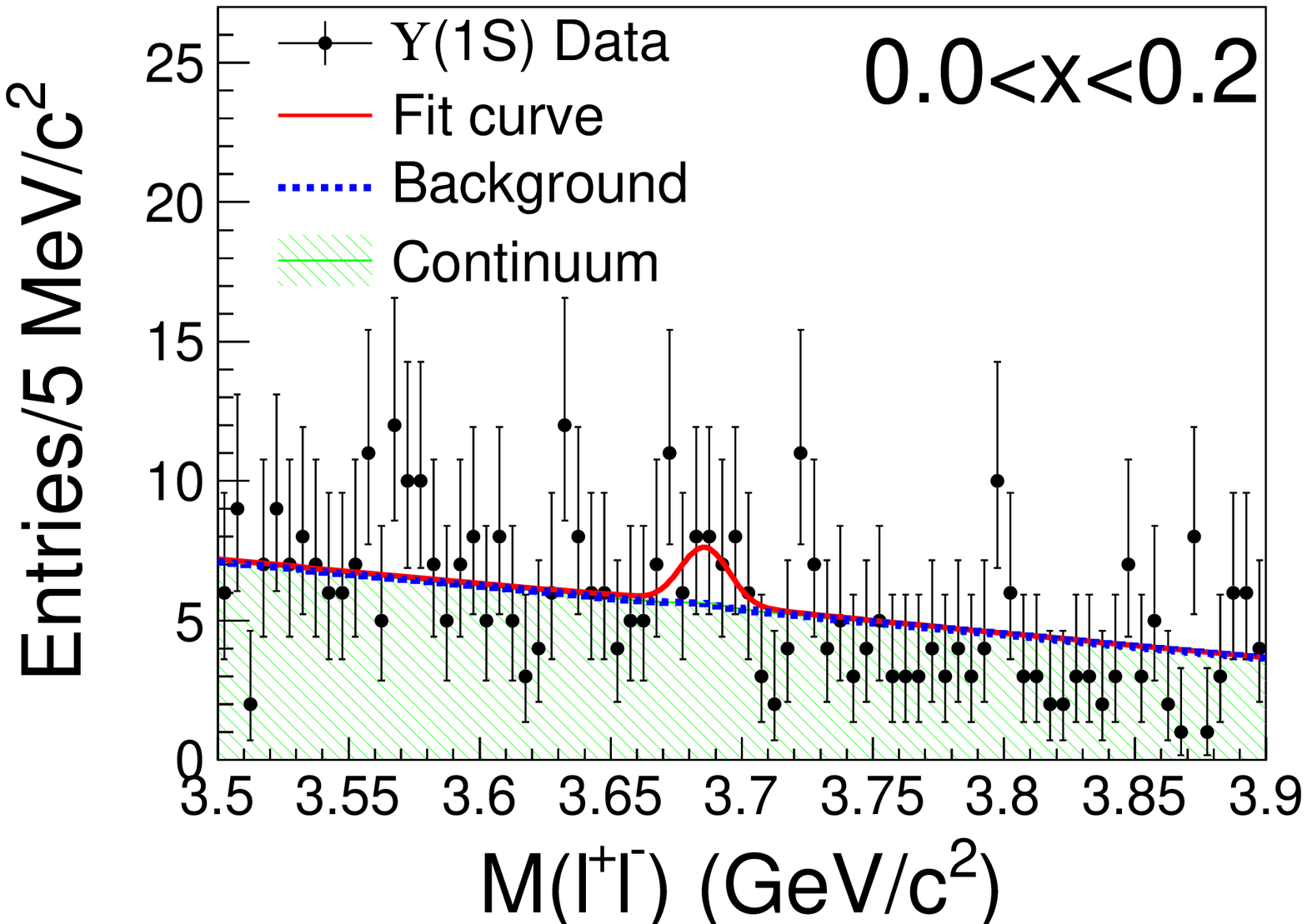}
    \includegraphics[width=0.3\textwidth]{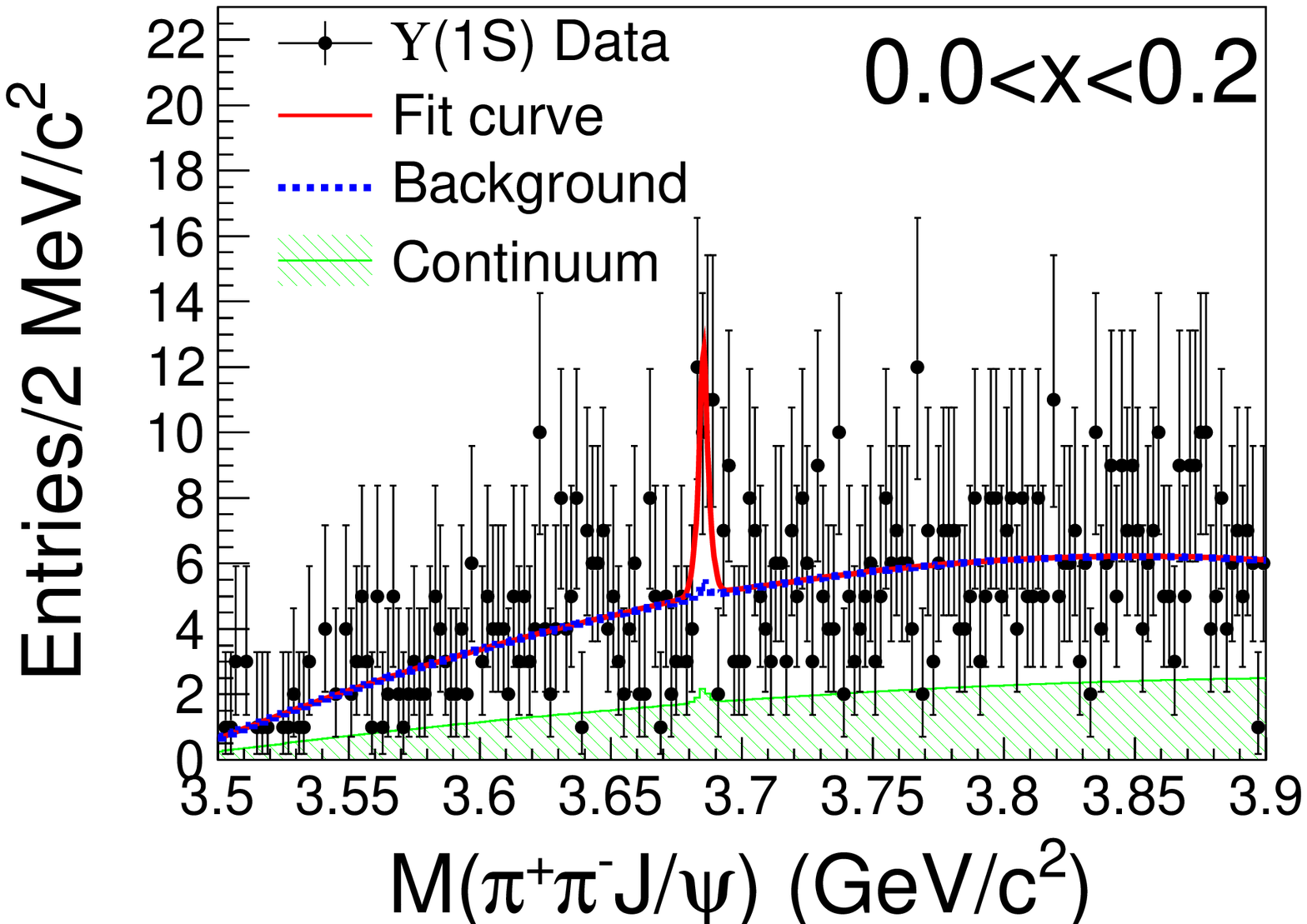}
    \includegraphics[width=0.3\textwidth]{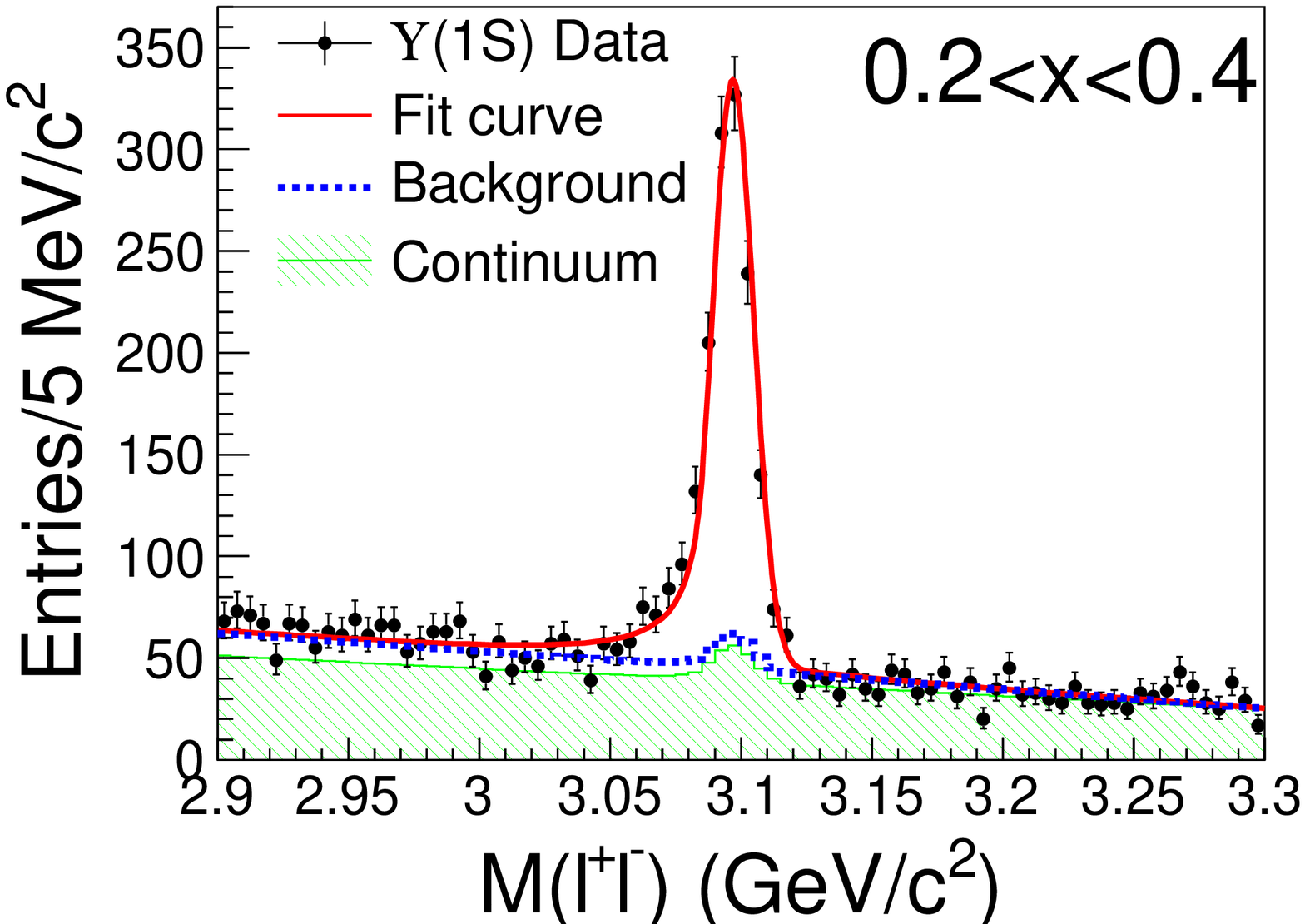}
    \includegraphics[width=0.3\textwidth]{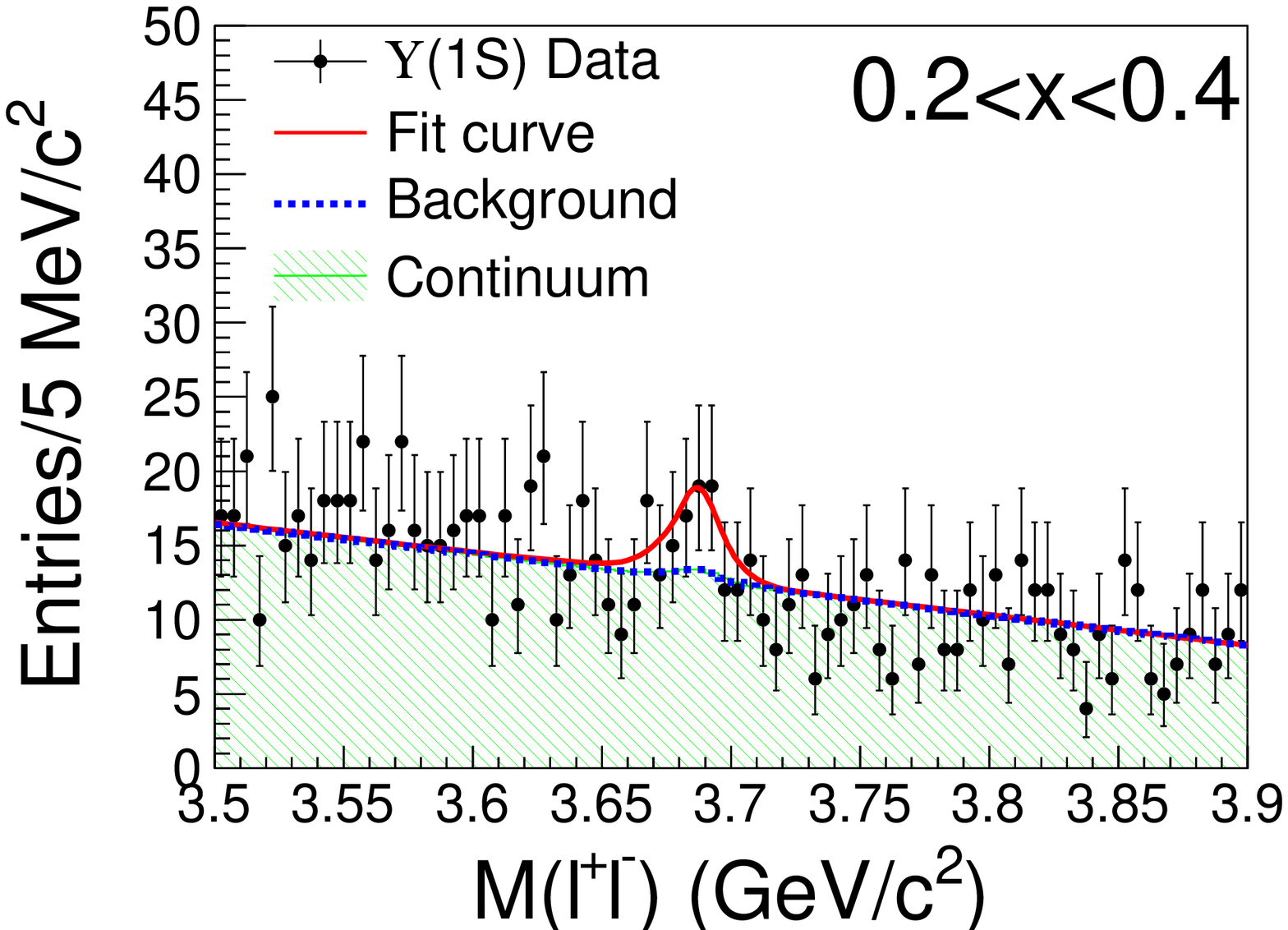}
    \includegraphics[width=0.3\textwidth]{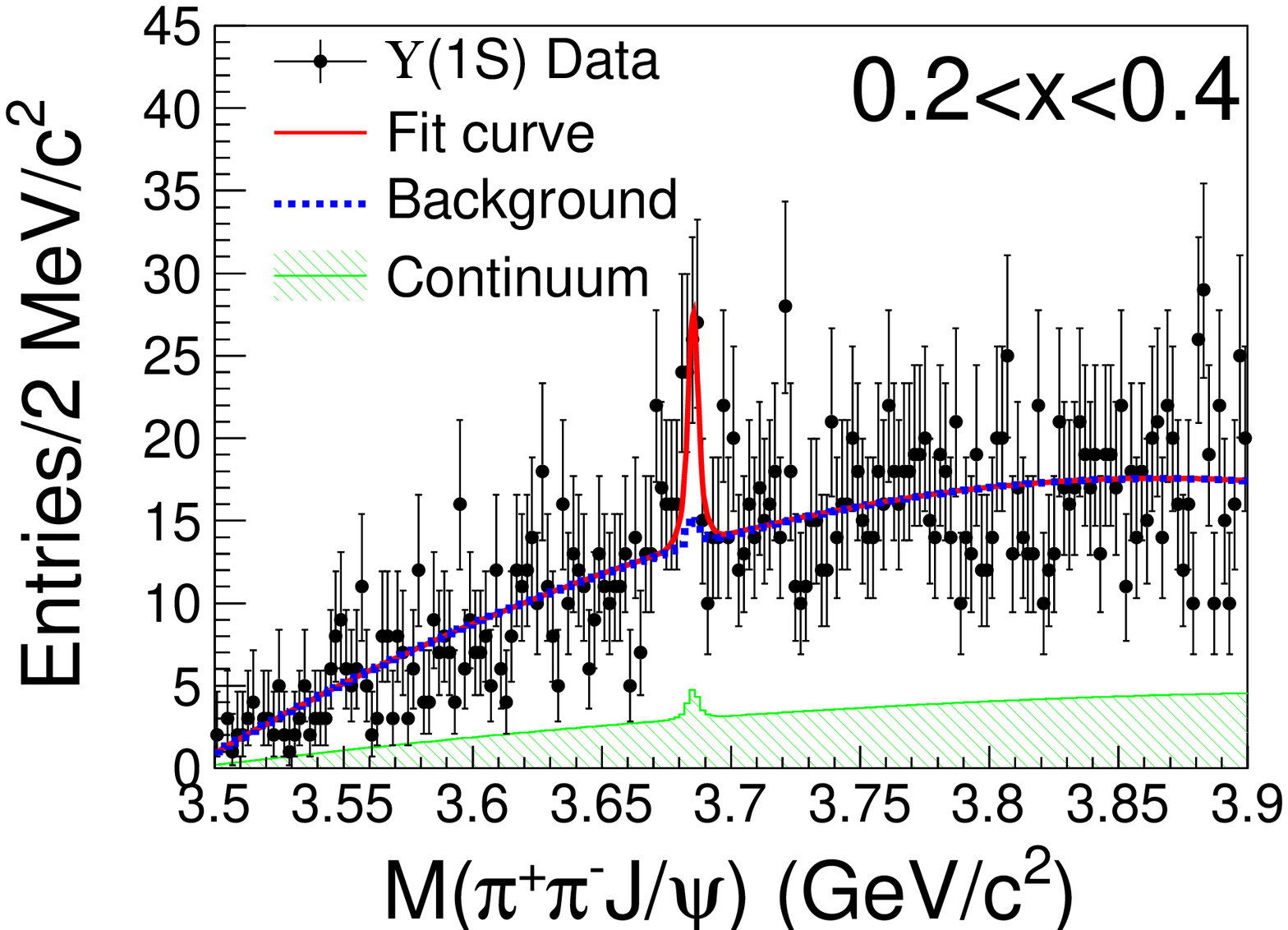}
    \includegraphics[width=0.3\textwidth]{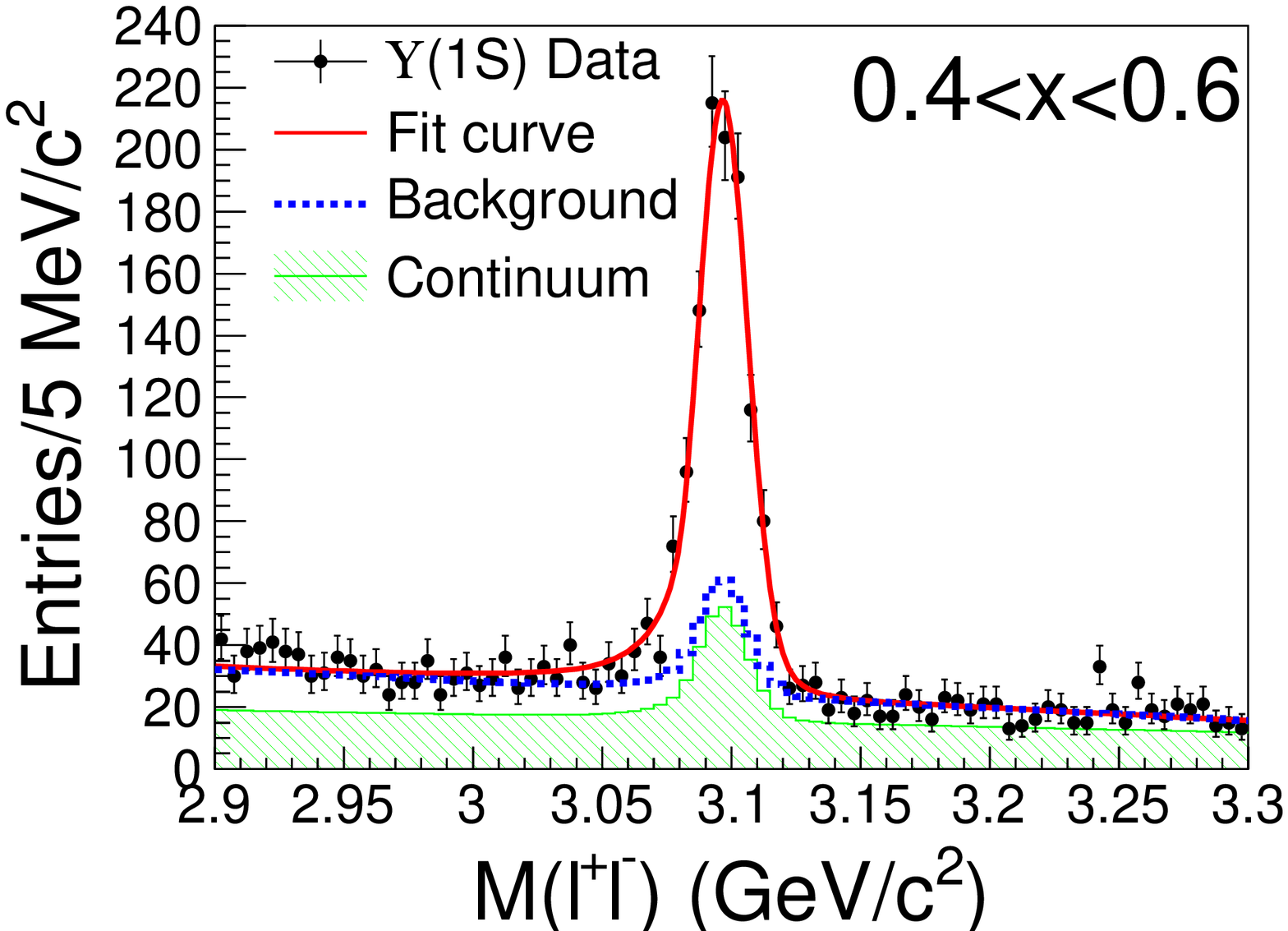}
    \includegraphics[width=0.3\textwidth]{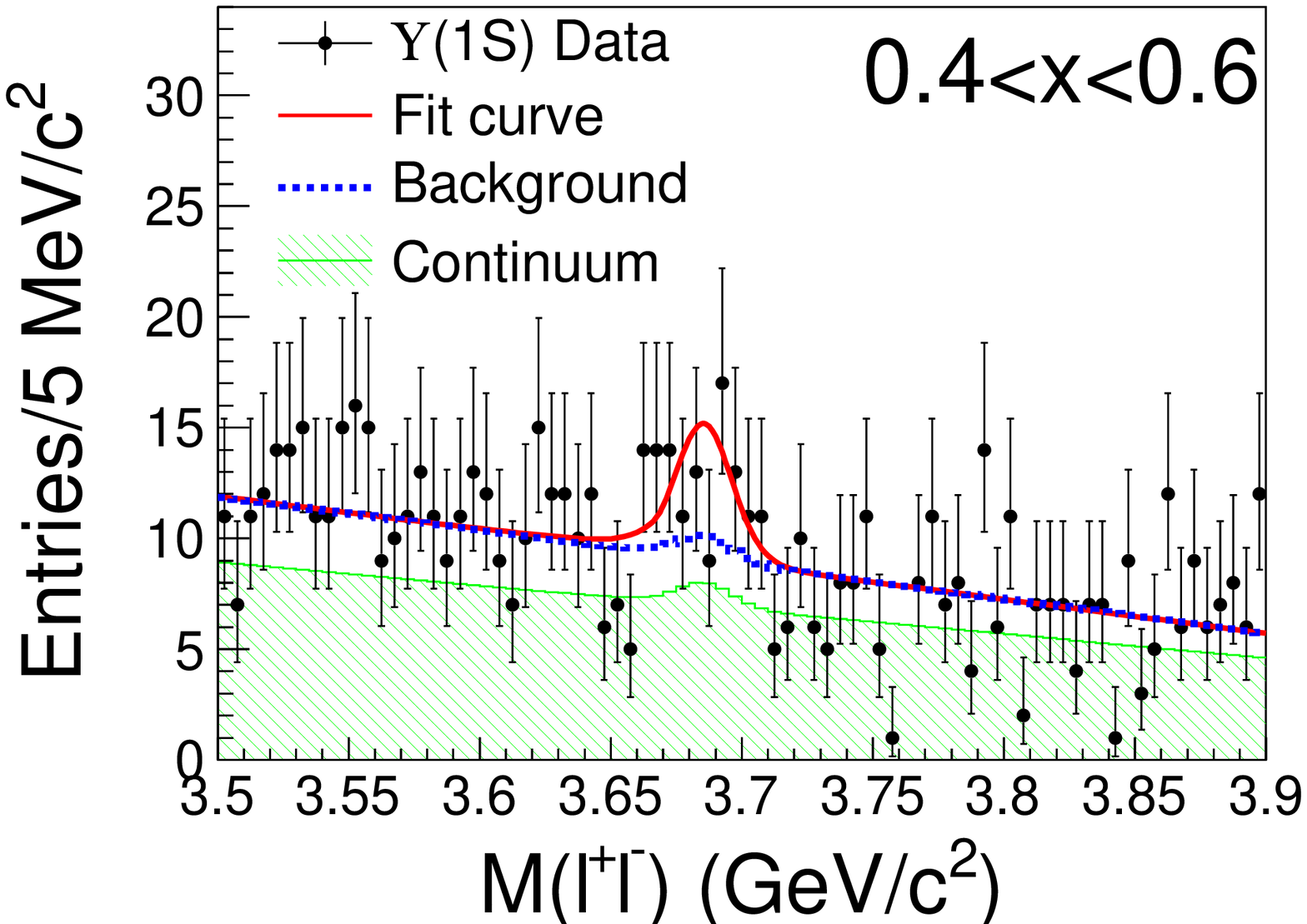}
    \includegraphics[width=0.3\textwidth]{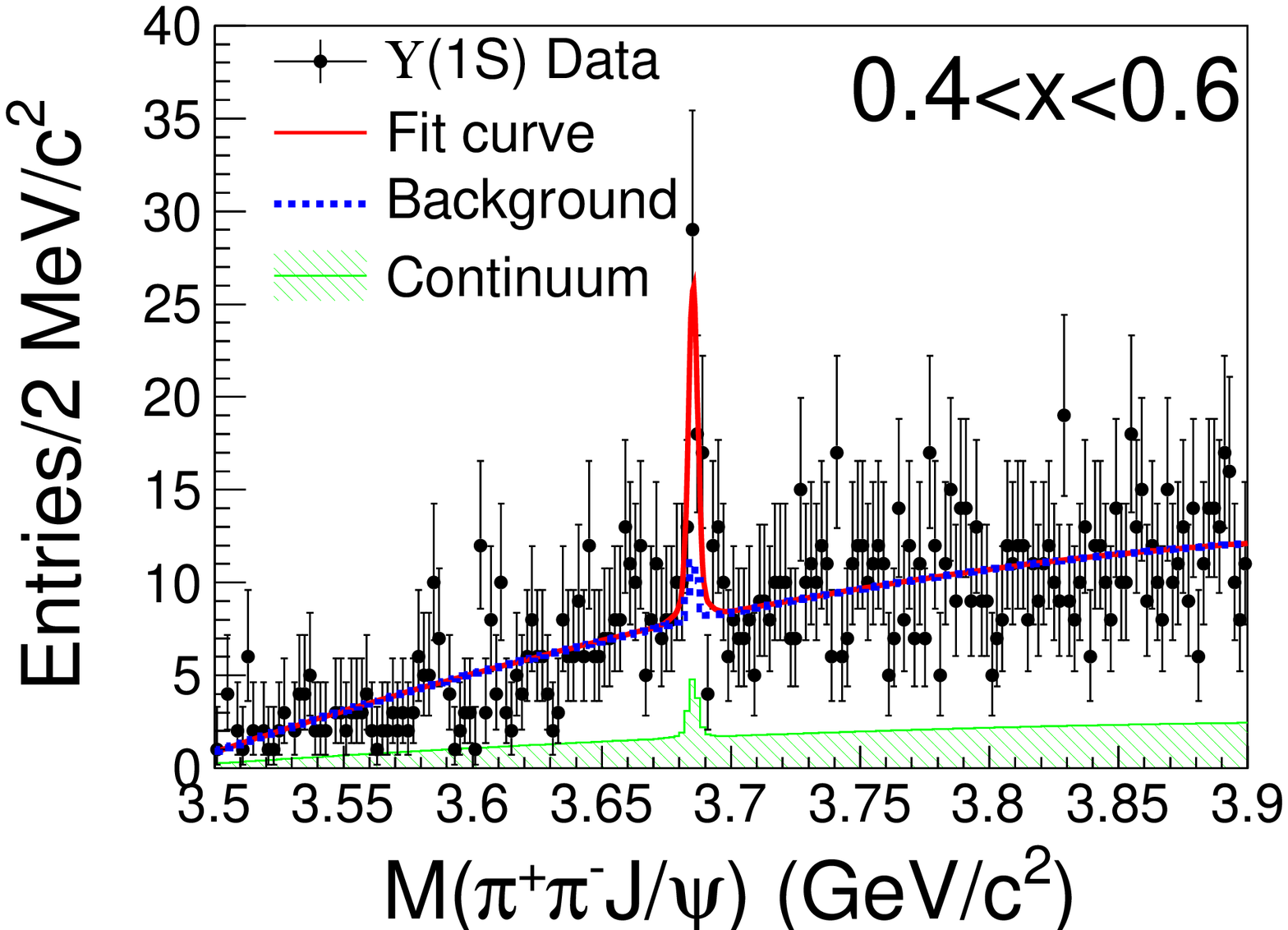}
    \includegraphics[width=0.3\textwidth]{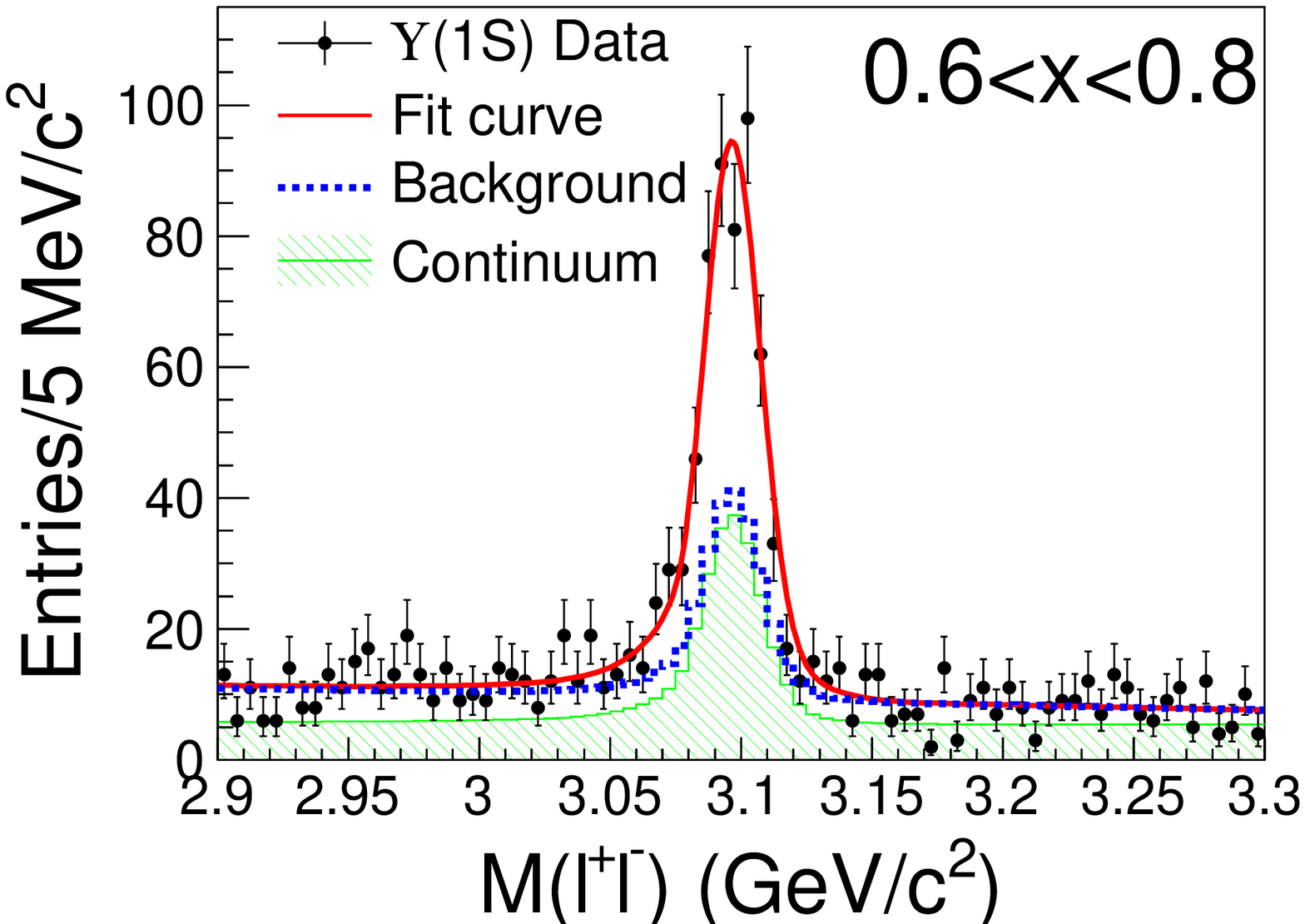}
    \includegraphics[width=0.3\textwidth]{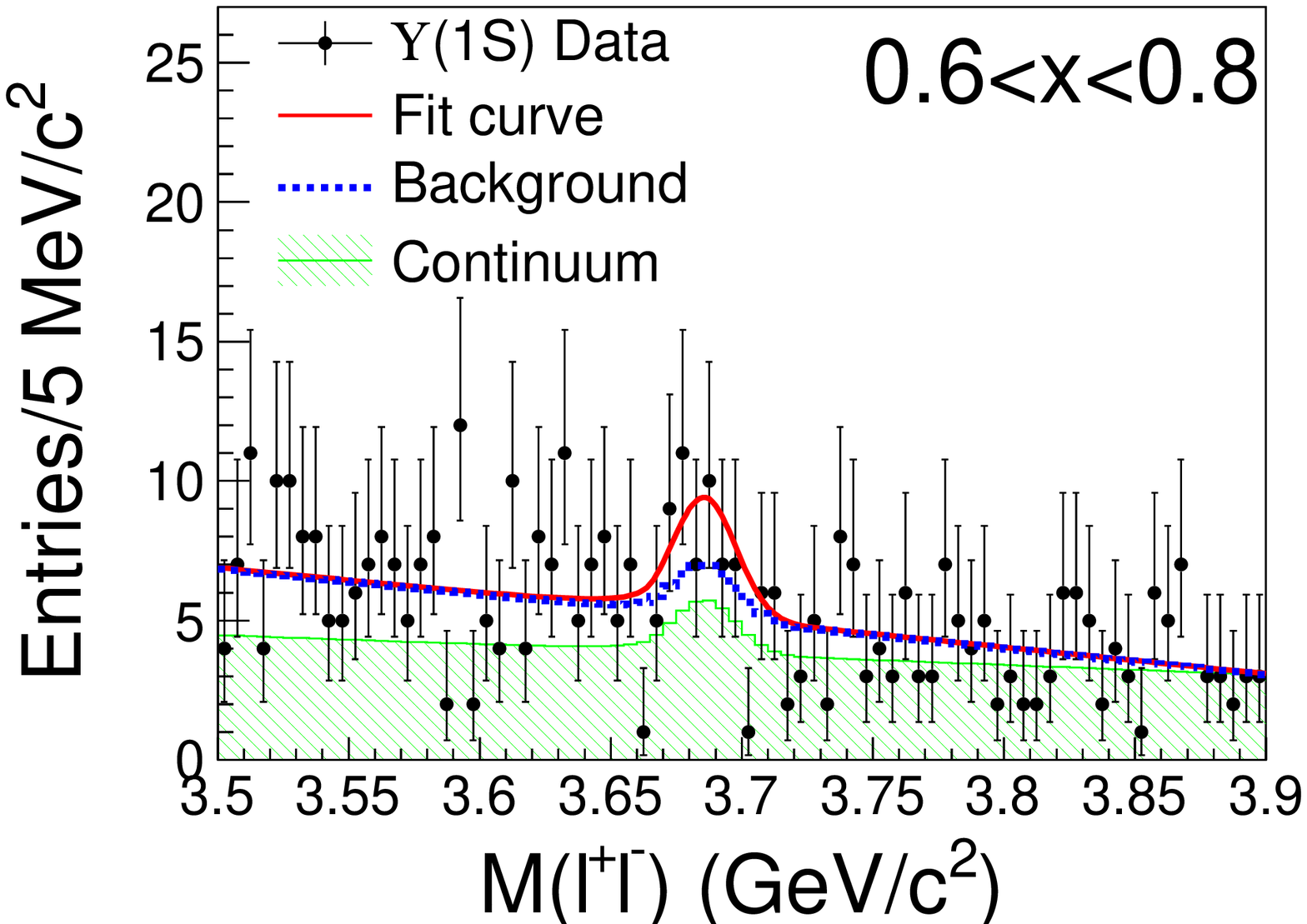}
    \includegraphics[width=0.3\textwidth]{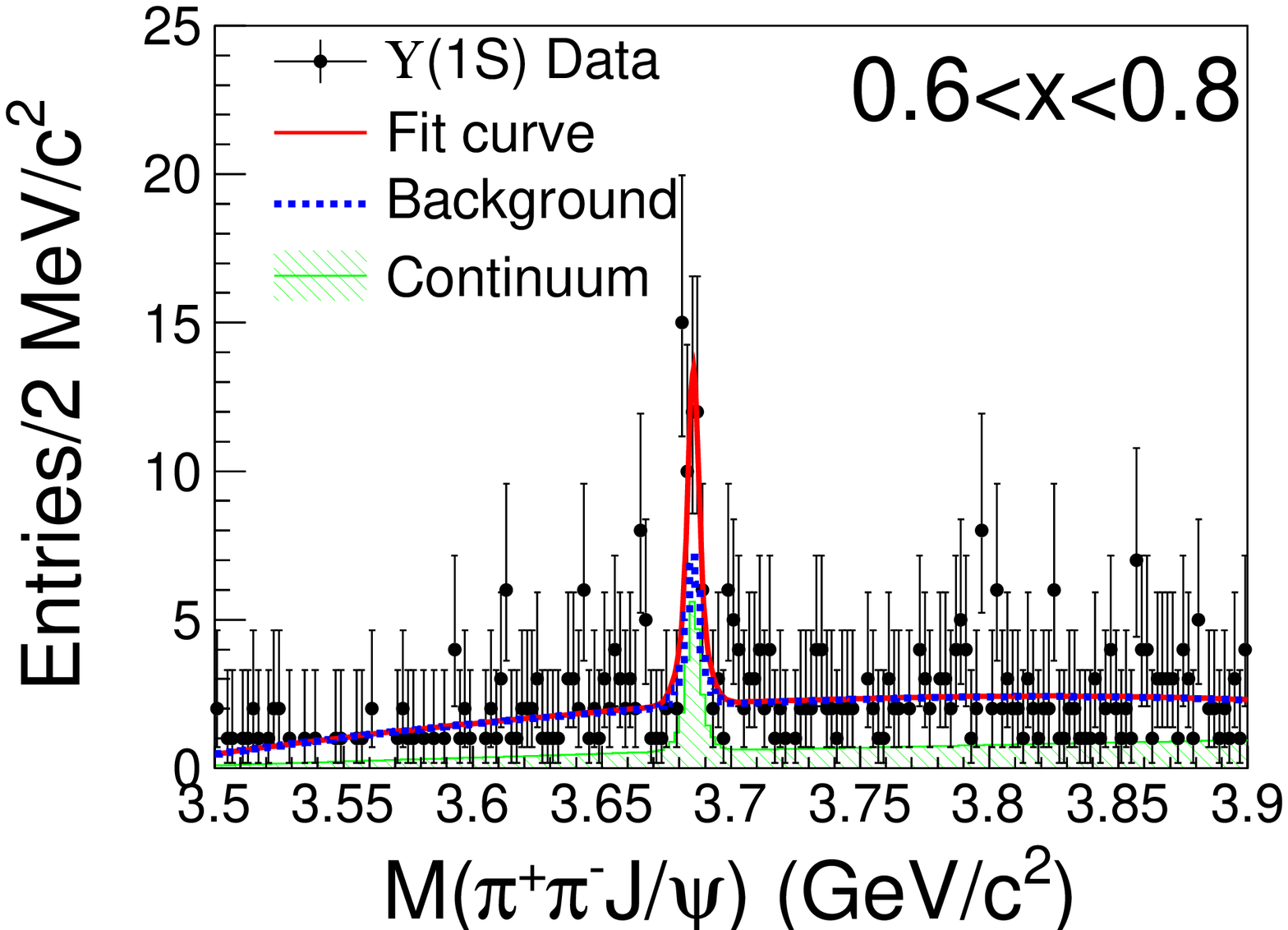}
    \includegraphics[width=0.3\textwidth]{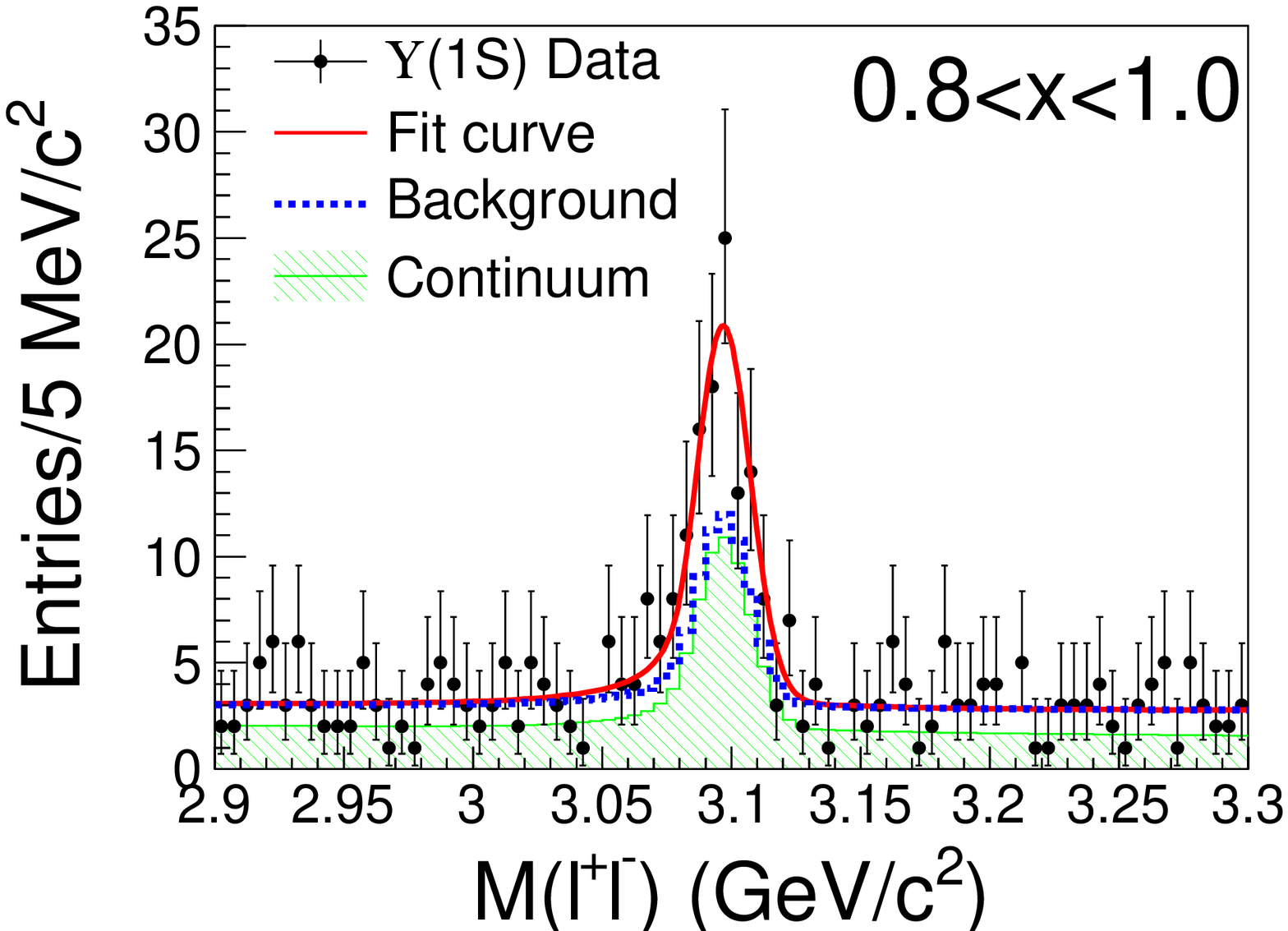}
    \includegraphics[width=0.3\textwidth]{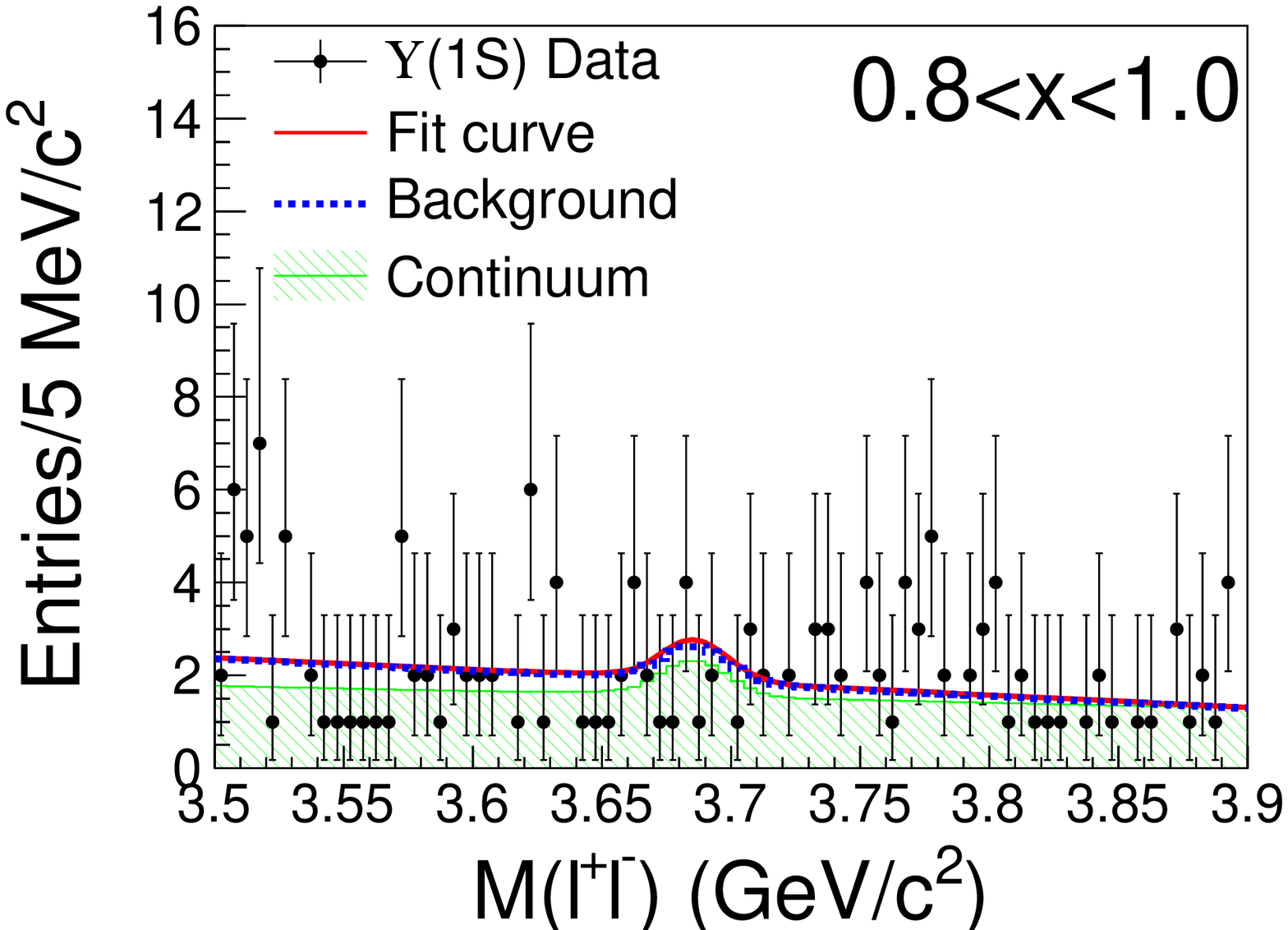}
    \includegraphics[width=0.3\textwidth]{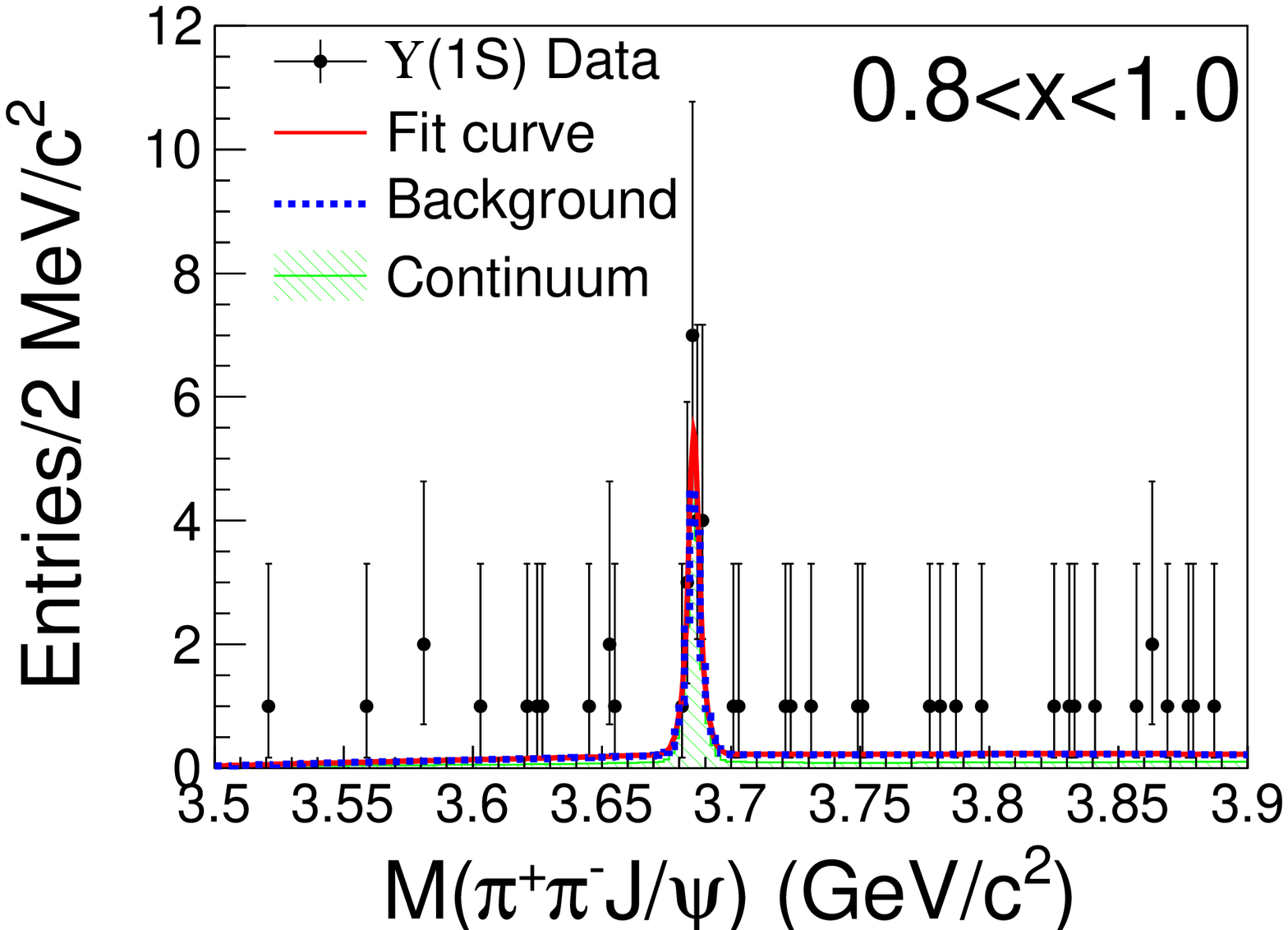}
  \caption{ Invariant mass distributions of the $J/\psi(\to\ell^+\ell^-)$ (left column),
  $\psi(2S)(\to\ell^+\ell^-)$ (middle column), and $\psi(2S)(\to\pi^+\pi^-J/\psi)$ (right column) candidates in
  the entire $x$ region (top row) and for $x$ bins of size $0.2$ (remaining rows).
  The points with error bars are for the $\Upsilon(1S)$ data sample;
  the shaded histograms are the continuum contributions scaled from the $\sqrt{s}=10.52~\mathrm{GeV}$ data sample.
  The solid lines are the best fit with the total fitted background components represented by the dashed lines.
  The $J/\psi$ and $\psi(2S)$ signal regions used for the $XYZ$ searches are indicated by the arrows in the top-row plots.}\label{fig-Psi-Y1S}
\end{figure*}

After all the event selection requirements,
significant $J/\psi(\to\ell^+\ell^-)$,
$\psi(2S)(\to\ell^+\ell^-)$, and $\psi(2S)(\to\pi^+\pi^-J/\psi)$ 
signals are seen in the
$\Upsilon(1S)$ data sample, as shown in
Fig.~\ref{fig-Psi-Y1S}.
The shaded histograms in this figure are the normalized continuum backgrounds that are determined
from the $\sqrt{s}=10.52~\mathrm{GeV}$ continuum data sample and
extrapolated down to the $\Upsilon(1S)$ resonance energy. The
scale factor used for this extrapolation is
$f_{{\rm scale}}=\mathcal{L}_{\Upsilon}/\mathcal{L}_{\rm con}\times \sigma_{\Upsilon}/\sigma_{\rm con}\times \varepsilon_{\Upsilon}/\varepsilon_{\rm con}$,
where $\mathcal{L}_{\Upsilon}/\mathcal{L}_{\rm con}$,
$\sigma_{\Upsilon}/\sigma_{\rm con}$, and
$\varepsilon_{\Upsilon}/\varepsilon_{\rm con}$ are the ratios
of the integrated luminosities, cross sections, and efficiencies,
respectively, for the $\Upsilon(1S)$ and continuum samples. The
MC-determined efficiencies for the $\Upsilon(1S)$ and continuum
data samples are found to be nearly the same for all the decay
modes, and the dependence of the cross sections on $s$ is
assumed to be proportional to
$1/s^2$~\cite{PhysRevD.69.094027,PhysRevD.56.321,hep-hp13108597}.
The resulting scale factor is 0.098.

Considering the slight differences in the MC-determined
reconstruction efficiencies for different $J/\psi(\psi(2S))$
momenta, we partition the data samples 
according to the scaled momentum
$x=p^{\ast}_{\psi}/(\frac{1}{2\sqrt{s}}\times(s-m_{\psi}^2))$~\cite{PhysRevD.70.072001},
where the subscript $\psi$ represents $J/\psi$ ($\psi(2S)$),
$p^{\ast}_{\psi}$ is the momentum of the $\psi$ candidate in the
$e^+e^-$ center-of-mass system, and $m_{\psi}$ is the $\psi$
mass~\cite{ChinPhysC38.090001}. The value of
$(\frac{1}{2\sqrt{s}}\times(s-m_{\psi}^2))$ is the value of
$p^{\ast}_{\psi}$ for the case where the $\psi$ candidate recoils
against a massless particle. The use of $x$ removes the
beam-energy dependence in comparing the continuum
data to that taken at the $\Upsilon(1S)$ resonance.

An unbinned extended simultaneous likelihood fit is applied to the
$x$-dependent $J/\psi(\psi(2S))$ spectra to extract the signal
yields in the $\Upsilon(1S)$ and continuum data samples. Due to
the slight dependence on momentum, the $J/\psi(\psi(2S))$
signal shape is directly obtained from the MC simulation in each
$x$ bin convolved with a Gaussian function with a free width
in the fit to account for possible discrepancy between data and MC simulation.
In the fit to the $\Upsilon(1S)$ candidates,
a Chebyshev polynomial background shape is used for the $\Upsilon(1S)$
 decay backgrounds in addition to the normalized continuum contribution.
Particularly for the
$\Upsilon(1S)$ to $\psi(2S)$ inclusive decays, the
$\psi(2S)\to\ell^{+}\ell^{-}$ and
$\psi(2S)\to\pi^{+}\pi^{-}J/\psi$ decay modes are treated together
to obtain the total $\psi(2S)$ signal yield; that is to say, we
apply an additional simultaneous fit to the $\psi(2S)$ candidates
in the two decay modes with the fixed ratios of MC-determined
efficiencies between them
with all of the branching fractions of the intermediate states included.

\begin{figure}[!htpb]
    %\centering
    \includegraphics[width=0.45\textwidth]{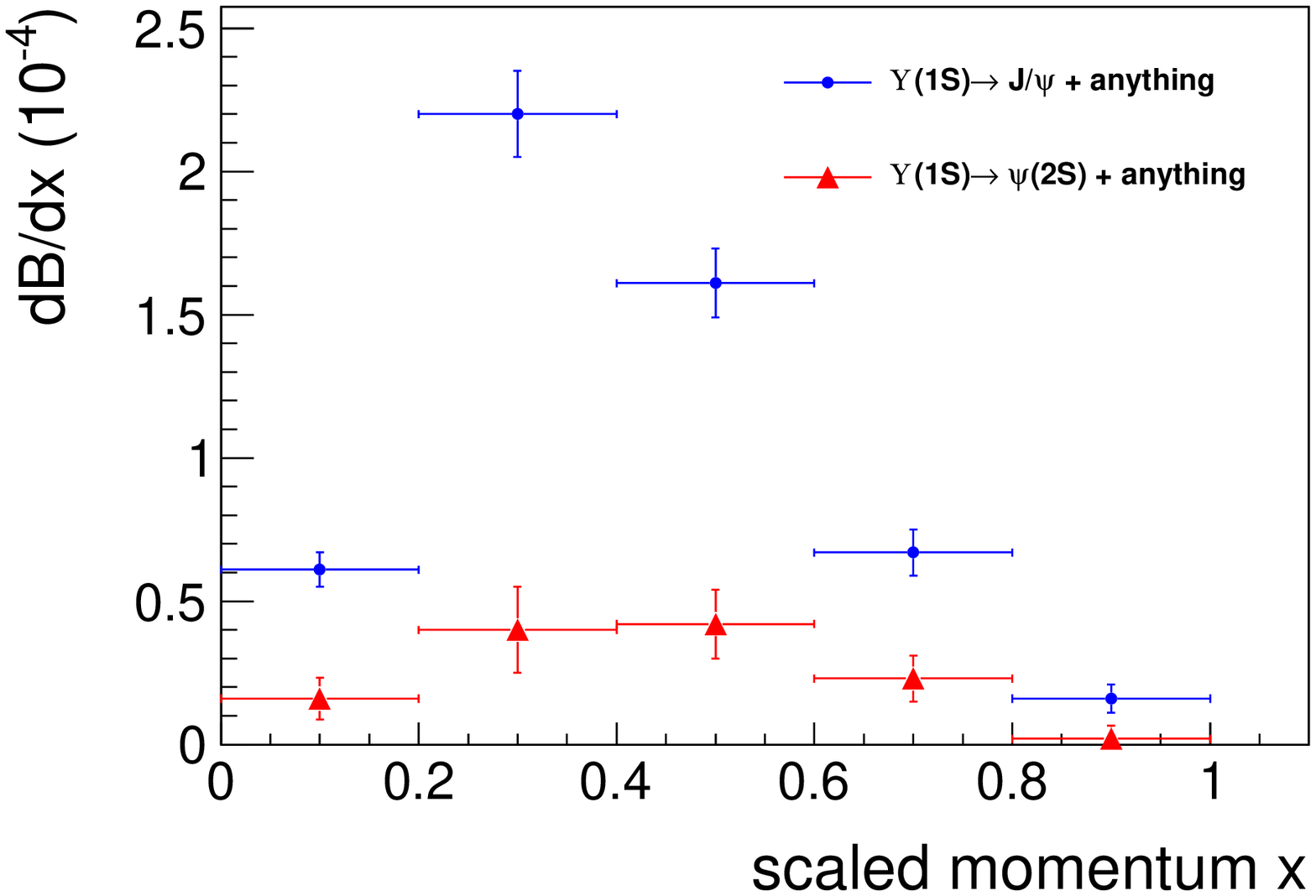}
  \caption{Differential branching fractions for $\Upsilon(1S)$ inclusive decays into  the $J/\psi$
  and $\psi(2S)$ versus the scaled momentum $x$ defined in the text. For each point,
  the error is the sum of the statistical and systematic errors.
  }\label{db}
\end{figure}

The invariant mass
distributions for the $J/\psi$ and $\psi(2S)$ candidates for the entire $x$ region
and $\Delta x = 0.2$ bins are shown in
Fig.~\ref{fig-Psi-Y1S}
with the results of the fits to the spectra of the $J/\psi$ and
$\psi(2S)$ candidates in $\Upsilon(1S)$ inclusive decays.
The fitted signal yields ($N_\mathrm{fit}$) in each $x$ bin are
tabulated in Table~\ref{Table-Nfit-Psi}, together with the
reconstruction efficiencies ($\varepsilon$) [including all intermediate-state
branching fractions], the total systematic
uncertainties ($\sigma_{\rm syst}$), and the
corresponding branching fractions ($\mathcal{B}$).
The total systematic uncertainties
are the sum of the common systematic errors (described
below) and fit errors estimated in each $x$ bin or 
the full range in $x$. The total numbers of  $J/\psi(\psi(2S))$ events, \textit{i.e.,} the sums of the
signal yields in all of the $x$ bins, the sums of the
$x$-dependent efficiencies weighted by the signal fraction in that
$x$ bin, and the measured branching fraction values are 
also itemized in Table~\ref{Table-Nfit-Psi}. Our measurements are consistent with the
PDG averages of previous results from
CLEO-c, but with
smaller central values and better precision.
In addition, Fig.~\ref{db} shows  the differential branching fractions of
$\Upsilon(1S)$ inclusive decays into  the $J/\psi$
and $\psi(2S)$.

\begin{table*}[t]
  \begin{threeparttable}
  \caption{\label{Table-Nfit-Psi} Summary of the branching fraction measurements of
  $\Upsilon(1S)$ inclusive decays into the $J/\psi(\psi(2S))$,
  where $N_{\rm fit}$ is the number of fitted signal events,
  $\varepsilon~(\%)$ is the reconstruction efficiency with all intermediate-state
branching fractions included,
  $\sigma_{\rm syst}(\%)$ is the total systematic error on the branching fraction measurement,
  and $\mathcal{B}$ is the measured branching fraction. For the $\psi(2S)$ channel, $\varepsilon$ is the sum of the reconstruction efficiencies
  in the $\ell^+\ell^-$ and $\pi^+\pi^-J/\psi$ decay modes
  with the branching fractions of the intermediate states included.}
  \begin{tabular}{c|r@{$\pm$}lccc||r@{$\pm$}lccc}
  \hline\hline
    \multicolumn{6}{c}{$\Upsilon(1S)\to J/\psi+{\rm anything}$} & \multicolumn{5}{c}{$\Upsilon(1S)\to\psi(2S)+{\rm anything}$} \\[-2pt]
   $x$ & \multicolumn{2}{c}{$N_{\rm fit}$} & $\varepsilon (\%)$ & $\sigma_{\rm syst} (\%)$ & $\mathcal{B}(10^{-4})$ & \multicolumn{2}{c}{$N_{\rm fit}$} & $\varepsilon (\%)$ & $\sigma_{\rm syst}(\%)$ & $\mathcal{B}(10^{-4})$ \\
  \hline
    $(0.0,0.2)$  & $379.3$  & $28.1$ & $6.06$  & $4.3$ & $0.61\pm0.05\pm0.03$ & $30.1$  & $10.5$ & $1.81$  & $21.8$ & $0.16\pm0.06\pm0.04$ \\
    $(0.2,0.4)$  & $1297.6$ & $48.6$ & $5.78$  & $5.4$ & $2.20\pm0.08\pm0.12$ & $71.3$  & $18.3$ & $1.76$  & $26.5$ & $0.40\pm0.10\pm0.11$ \\
    $(0.4,0.6)$  & $904.6$  & $41.6$ & $5.51$  & $5.6$ & $1.61\pm0.07\pm0.09$ & $71.5$  & $15.4$ & $1.68$  & $18.6$ & $0.42\pm0.09\pm0.08$ \\
    $(0.6,0.8)$  & $354.0$  & $29.3$ & $5.15$  & $6.8$ & $0.67\pm0.06\pm0.05$ & $39.5$  & $12.0$ & $1.65$  & $16.6$ & $0.23\pm0.07\pm0.04$ \\
    $(0.8,1.0)$  & $54.2$   & $13.4$ & $3.36$  & $7.6$ & $0.16\pm0.04\pm0.02$ & $2.5$   & $5.7$  & $1.40$  & $78.4$ & $0.02\pm0.04\pm0.02$ \\
    Sum          & $2989.6$ & $75.0$ & $5.62$  & $4.7$ & $5.25\pm0.13\pm0.25$ & $214.9$ & $29.3$ & $1.71$  & $8.9$ & $1.23\pm0.17\pm0.11$ \\
  \hline\hline
  \end{tabular}
  \end{threeparttable}
\end{table*}

We search for signals for certain $XYZ$ states by combining
the $J/\psi(\psi(2S))$ with one or two light charged hadrons
($K^\pm/\pi^\pm$). MC simulations indicate that the
mass resolutions of the $J/\psi(\psi(2S))$ candidates have a weak
dependence on the production mode, so common signal and sideband
regions are defined. In the $\phi J/\psi$ mode, the $\phi$
candidates are reconstructed in the $K^+K^-$ final state. For
$J/\psi$, $\psi(2S)$ and $\phi$ candidates in their decay
channels, the selected signal regions and the corresponding
sidebands are summarized in Table~\ref{region}. All sidebands
are defined to be twice as wide as the corresponding signal
region. No peaking backgrounds or evident structures are found in these
sideband events in any of the invariant mass distributions discussed below.
To improve the mass resolutions of $XYZ$
candidates, vertex and mass-constrained fits are applied to the $J/\psi(\psi(2S))$
candidates; an unconstrained-mass vertex fit is done for the $\phi$ candidates since
their natural width is larger than the mass resolution.

An unbinned extended simultaneous maximum likelihood fit to the mass distributions
of the $XYZ$ candidates is performed to extract the signal and background yields in the
$\Upsilon(1S)$ and continuum data samples. The signal shapes of the examined $XYZ$ states used
in the fits are obtained directly from MC simulations that use world
average values for their masses and
widths~\cite{ChinPhysC38.090001}.
In the fit to the $\Upsilon(1S)$ data sample, a Chebyshev
polynomial function is used for the $\Upsilon(1S)$
decay backgrounds in addition to the normalized continuum contribution.

\begin{table}[!htpb]
  \begin{threeparttable}
  \caption{\label{region} The definitions of the signal regions and the corresponding sidebands
   for (a) $J/\psi \to \ell^+\ell^-$, (b) $\psi(2S)\to\ell^{+}\ell^{-}$,
  (c) $\psi(2S)\to\pi^{+}\pi^{-}J/\psi$, and (d) $\phi \to K^+ K^-$.
  The sidebands are selected to be twice as wide as the corresponding signal region.}
  \begin{tabular}{ccc}
  \hline\hline
  Channel & Signal Region & Sidebands (GeV/$c^2$)  \\
  \hline
    (a)  & [3.067, 3.127] & [2.970, 3.030] or [3.170, 3.230]\\
    (b)  & [3.6485, 3.7235] & [3.535, 3.610] or [3.760, 3.835]  \\
    (c)  & [3.677, 3.695] & [3.652, 3.670] or [3.700, 3.718] \\
    (d) & [1.012, 1.027] & [0.989, 1.004] or [1.036, 1.051] \\
  \hline\hline
  \end{tabular}
  \end{threeparttable}
\end{table}

%%%%%%%%%%%%%%%%%%%%%%%%%%%%%%%%%%%%%%%%%%%%%%%%%%%%%%%%%%%%%%%%%%%%%%%%%%%%%%%%%%%%%%%%%%%%%%%%%%%
%%%%%%%%%%%%%%%%%%%%%% Y(1S)->XYZ+anything->J/psi(psi(2S))hh+anything %%%%%%%%%%%%%%%%%%%%%%%%%%%%%%%
%%%%%%%%%%%%%%%%%%%%%%%%%%%%%%%%%%%%%%%%%%%%%%%%%%%%%%%%%%%%%%%%%%%%%%%%%%%%%%%%%%%%%%%%%%%%%%%%%%%

\begin{figure*}[htbp]
\includegraphics[width=0.4\textwidth]{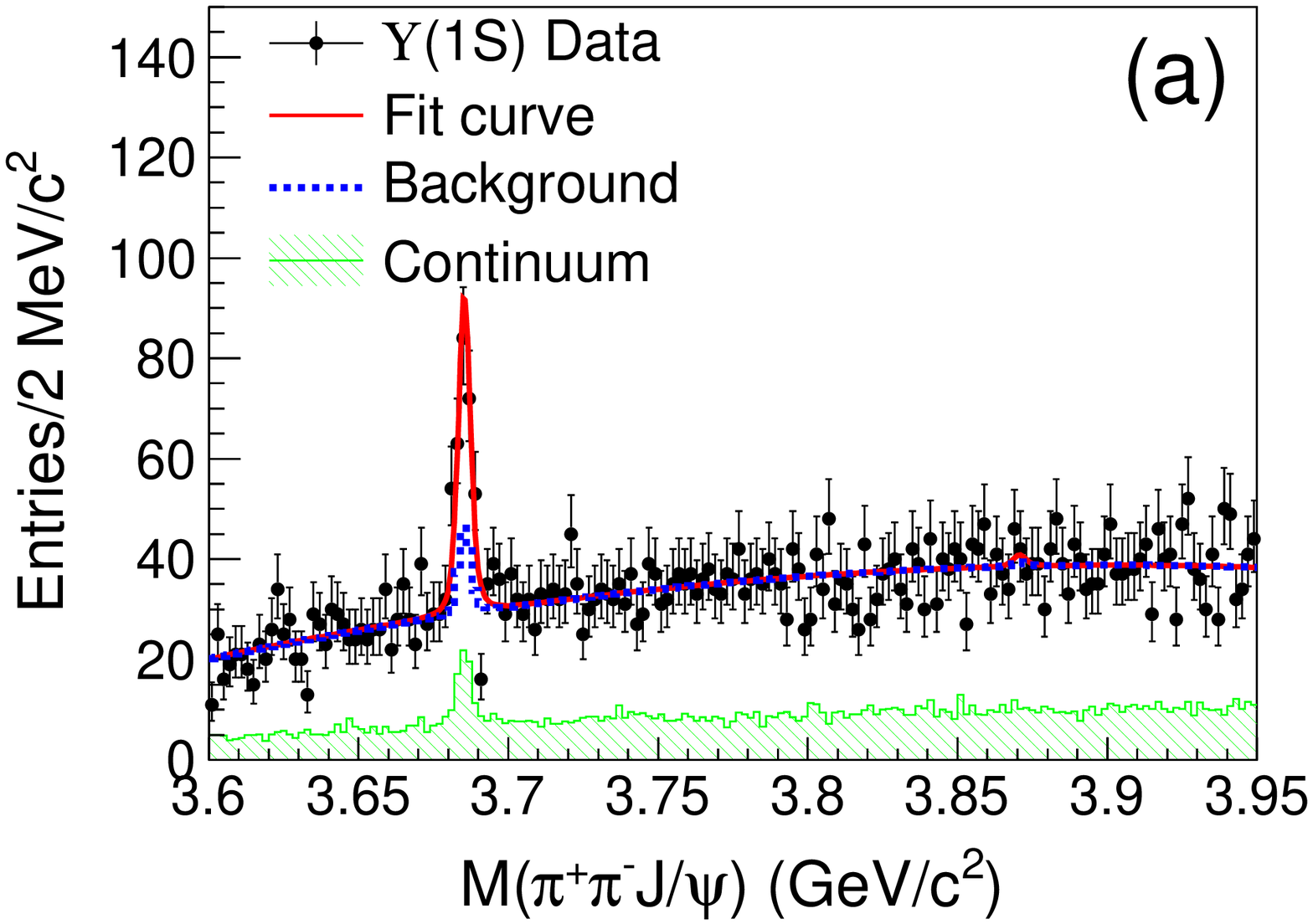}% \hspace{-0.25cm}
\includegraphics[width=0.4\textwidth]{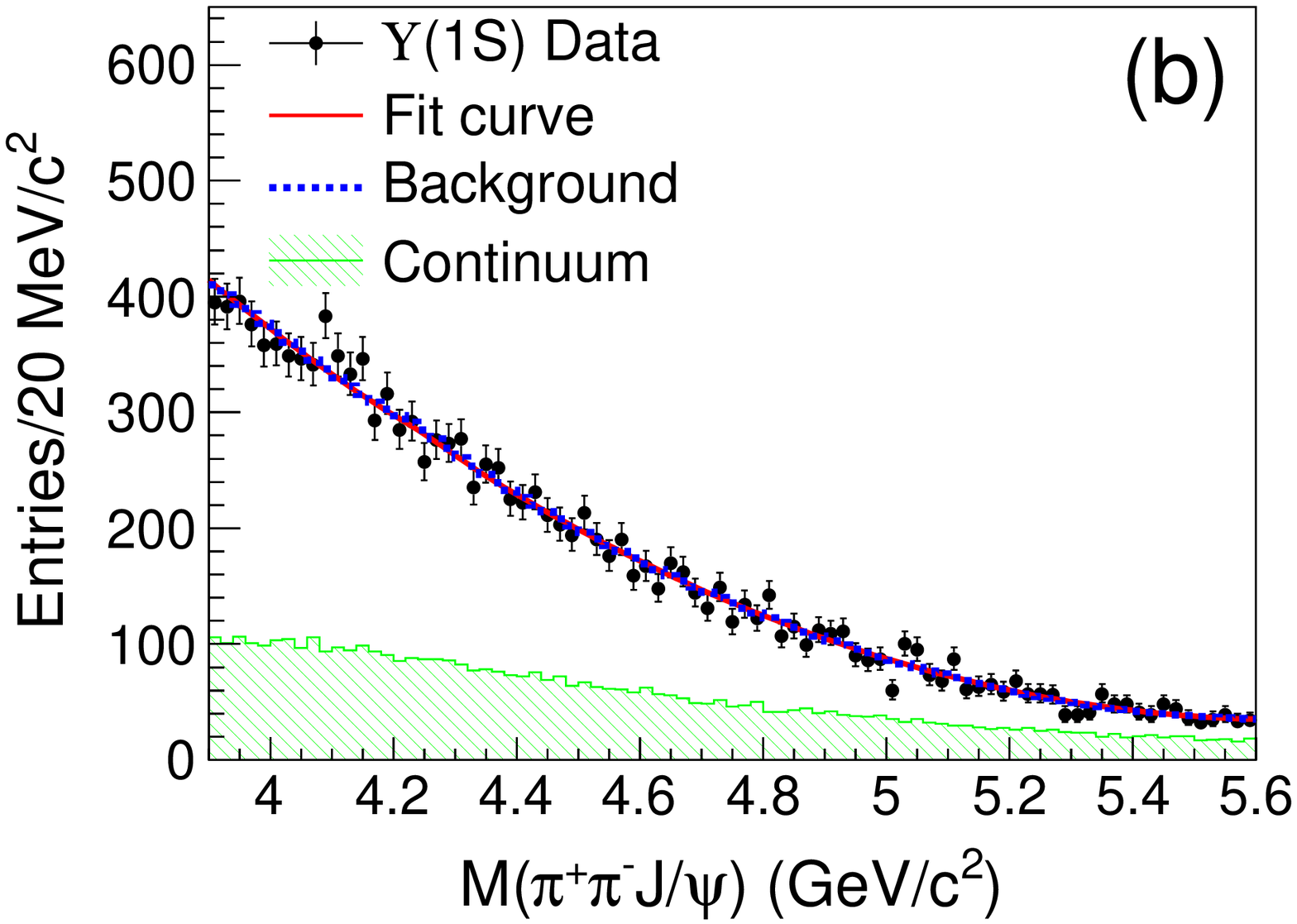}
\includegraphics[width=0.4\textwidth]{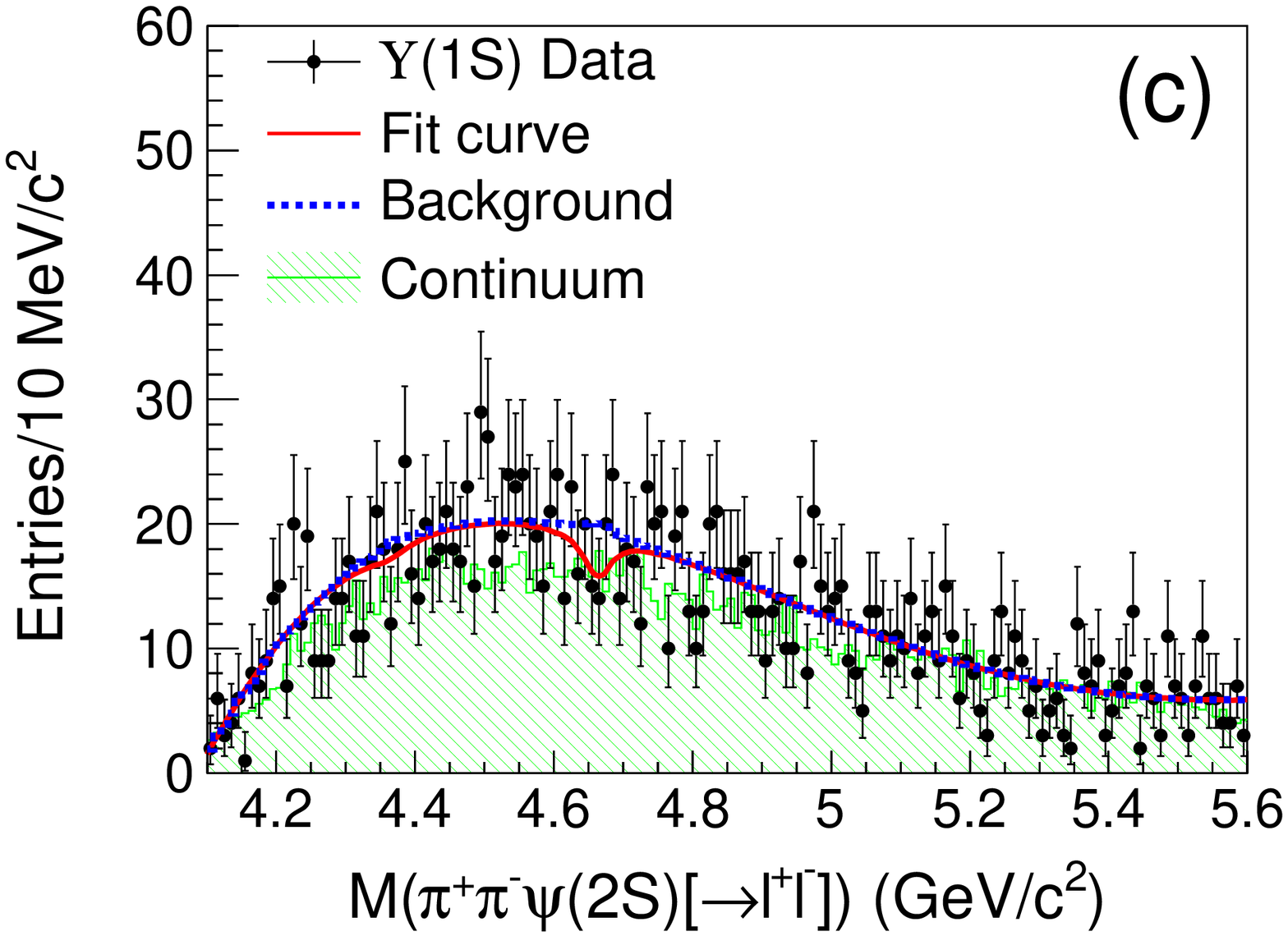}%\hspace{-0.25cm}
\includegraphics[width=0.4\textwidth]{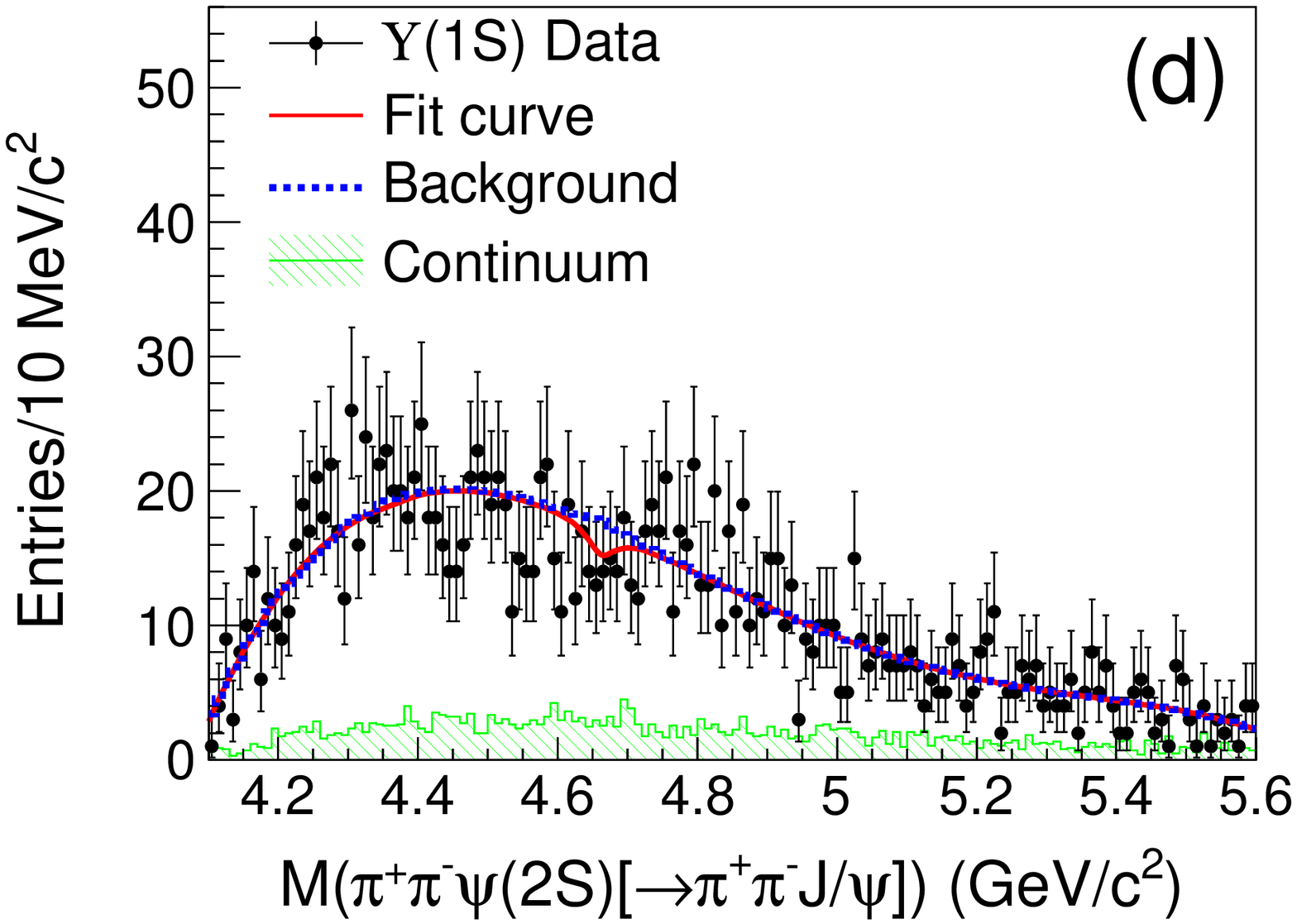}
  \caption{The $\pi^+\pi^-J/\psi$ invariant mass distributions for the (a) lower- and (b)
  higher-mass regions;  the (c) $\pi^+\pi^-\psi(2S)(\to\ell^+\ell^-)$ and
  (d) $\pi^+\pi^-\psi(2S)(\to\pi^+\pi^-J/\psi)$  invariant
  mass distributions.
  The points with error bars are the $\Upsilon(1S)$ events
  and the shaded histograms are the scaled continuum contributions
  determined from the data sample collected at $\sqrt{s}=10.52~\mathrm{GeV}$.
  The solid lines are the best fits with the total background components
  represented by the dashed lines.}\label{fig-X-Y2X3872}
\end{figure*}

\begin{figure*}[htbp]
\includegraphics[width=0.34\textwidth]{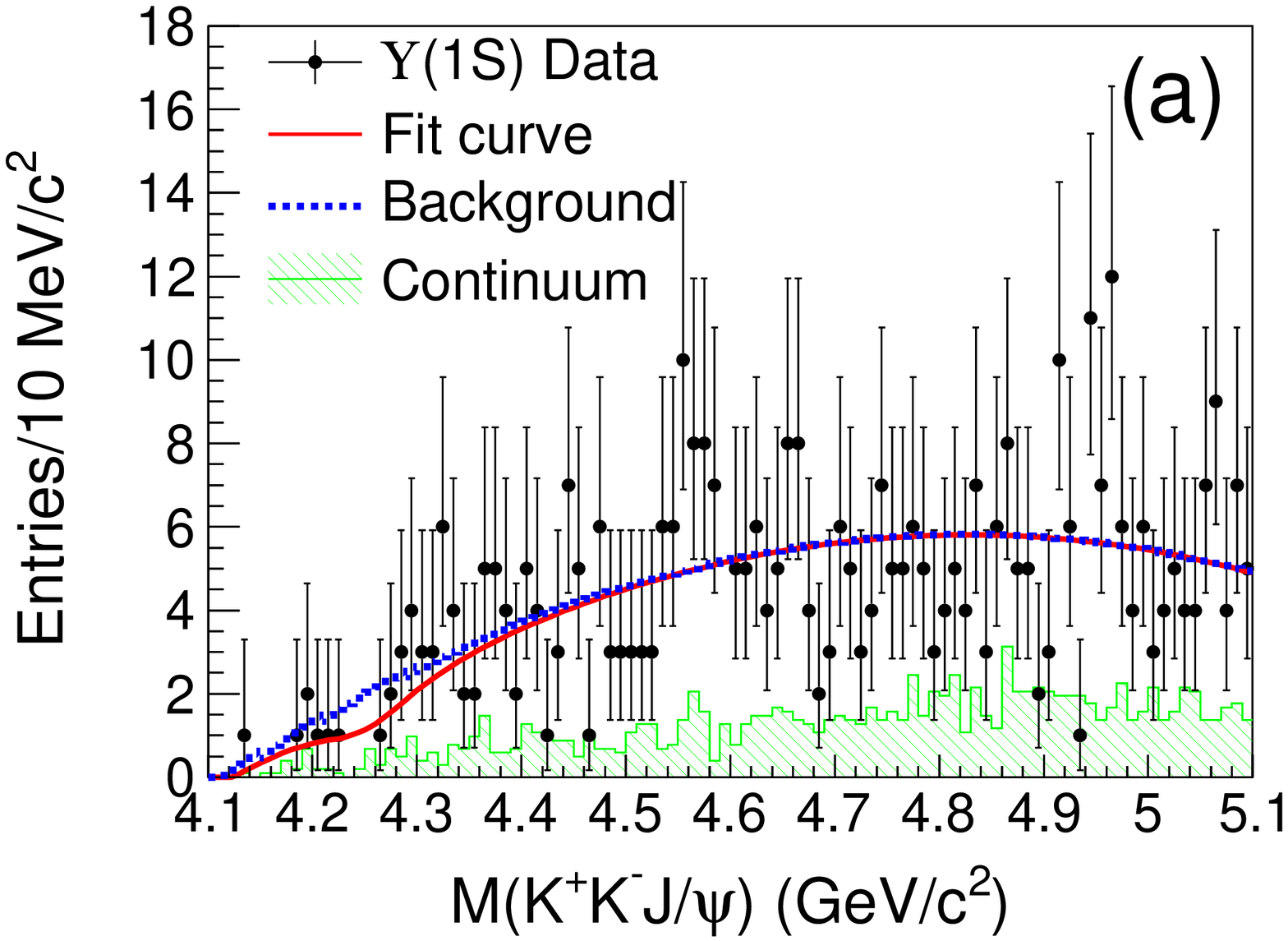} \hspace{-0.35cm}
\includegraphics[width=0.34\textwidth]{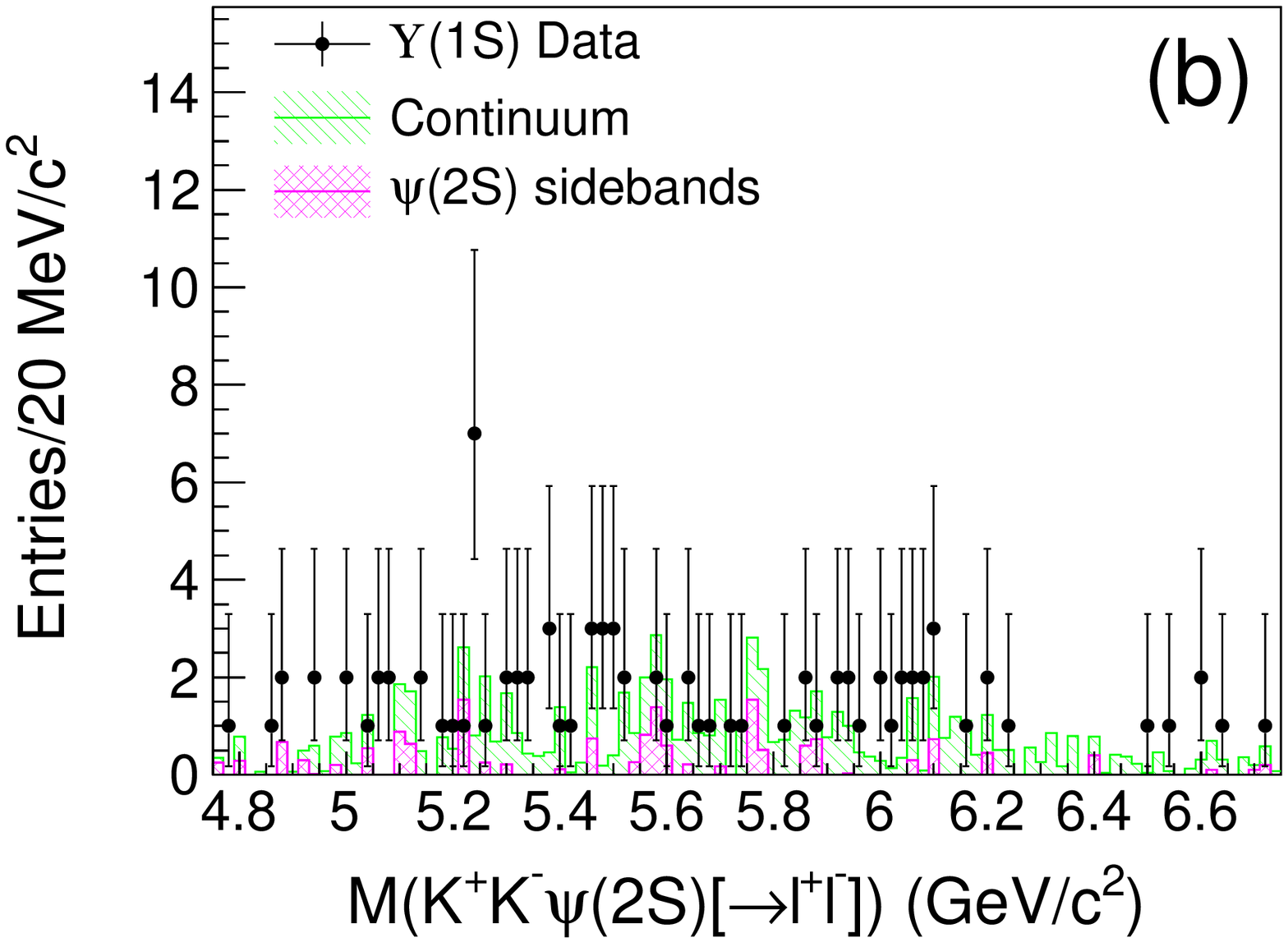} \hspace{-0.35cm}
\includegraphics[width=0.34\textwidth]{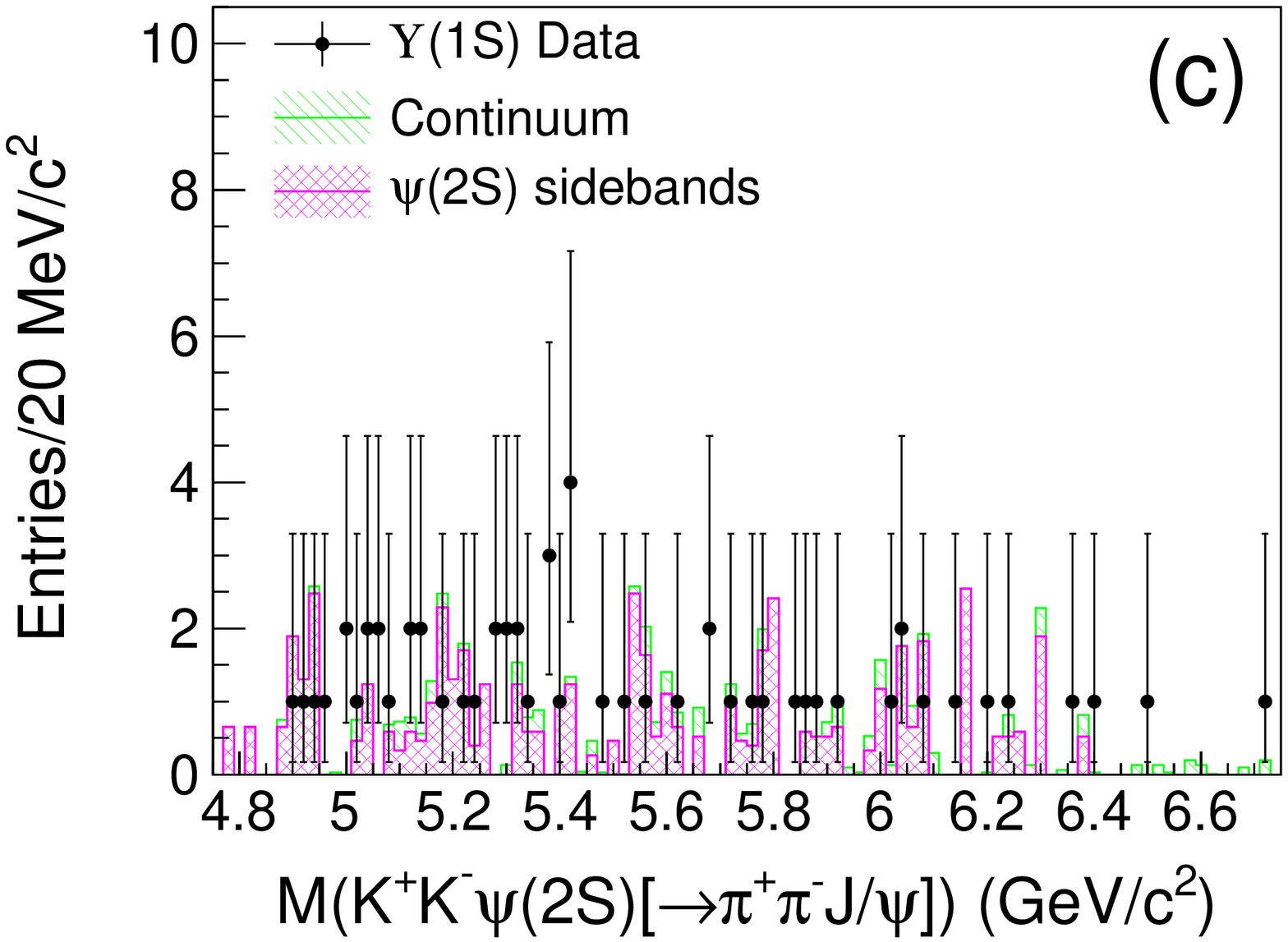}
  \caption{
  Invariant mass distributions of the (a) $K^+K^- J/\psi$,
  (b) $K^+K^- \psi(2S)(\to\ell^+\ell^-)$, and (c) $K^+K^- \psi(2S)(\to\pi^+\pi^-J/\psi)$ candidates
  in $\Upsilon(1S)$ inclusive decays.
  The points with error bars are the $\Upsilon(1S)$ events
  and the slant-shaded histograms are the scaled continuum
  contributions with the data sample collected at $\sqrt{s}=10.52~\mathrm{GeV}$
  which overlie the normalized $\psi(2S)$ mass sideband backgrounds
  (the cross-shaded histograms) for the two $\psi(2S)$ decay modes.
  The solid line in panel (a) is the best fit with the fitted total background component represented as
  a dashed line.}\label{fig-X-KKJpsi}
\end{figure*}

\begin{figure*}[htbp]
\includegraphics[width=0.34\textwidth]{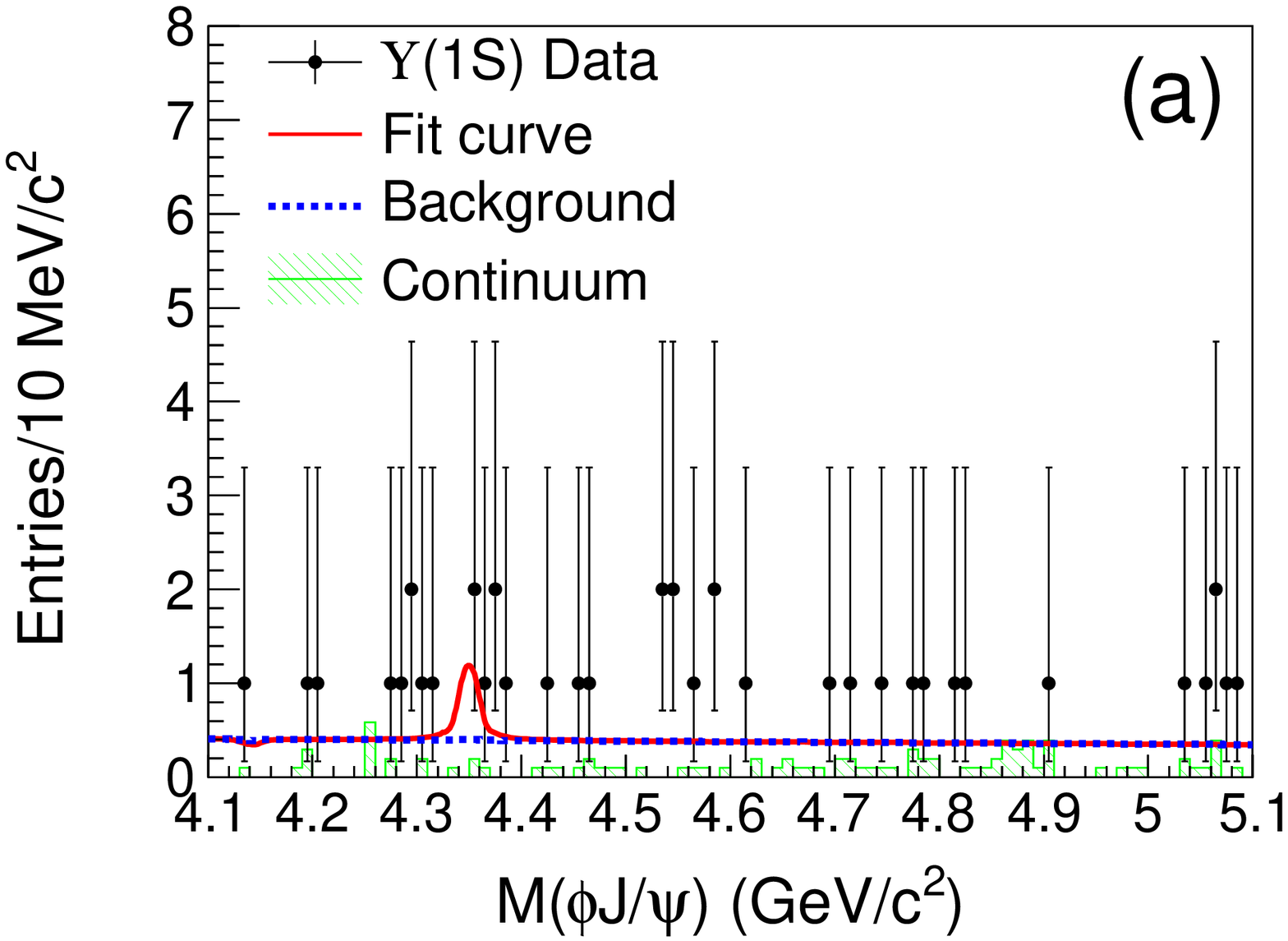} \hspace{-0.35cm}
\includegraphics[width=0.34\textwidth]{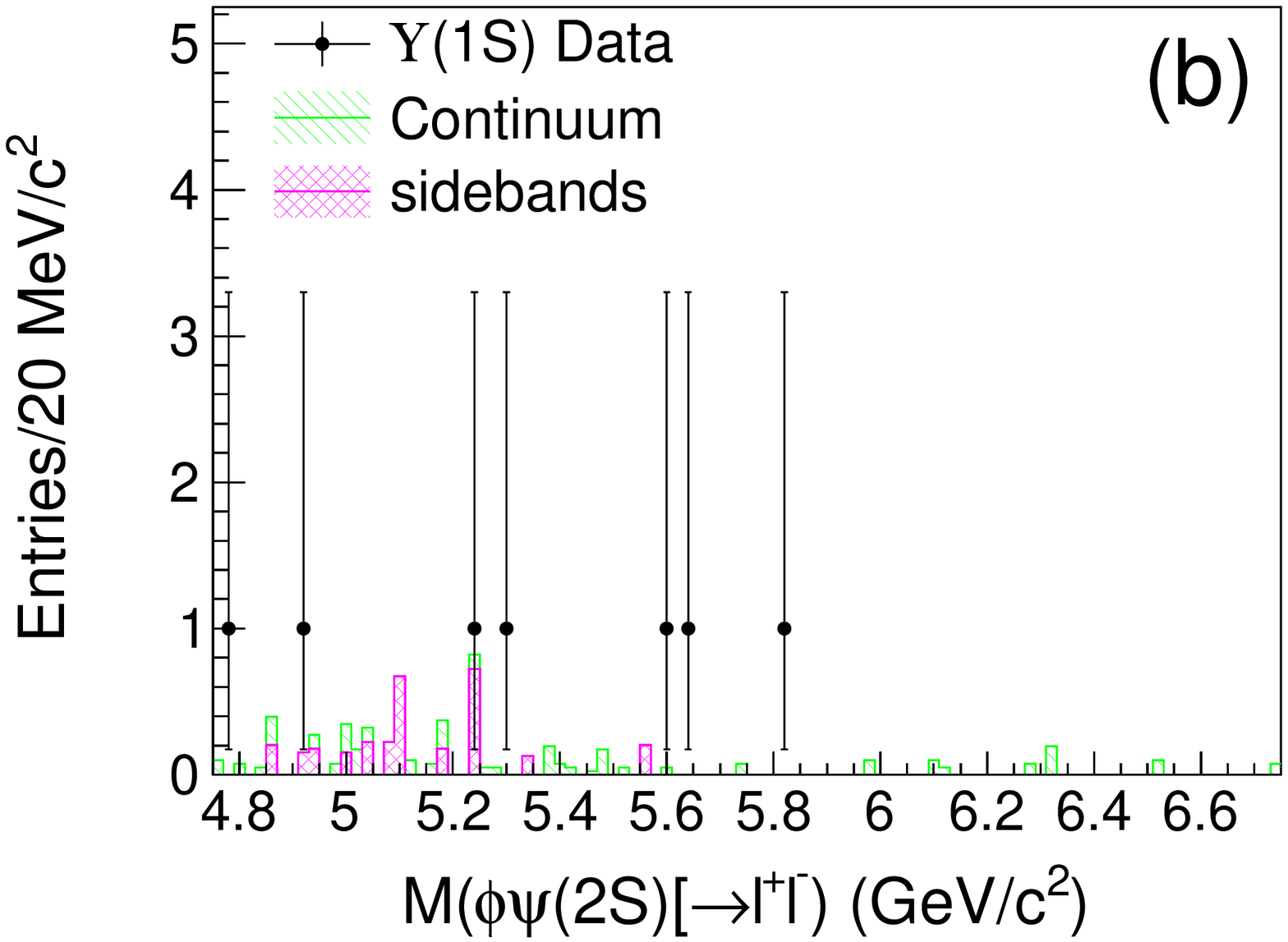} \hspace{-0.35cm}
\includegraphics[width=0.34\textwidth]{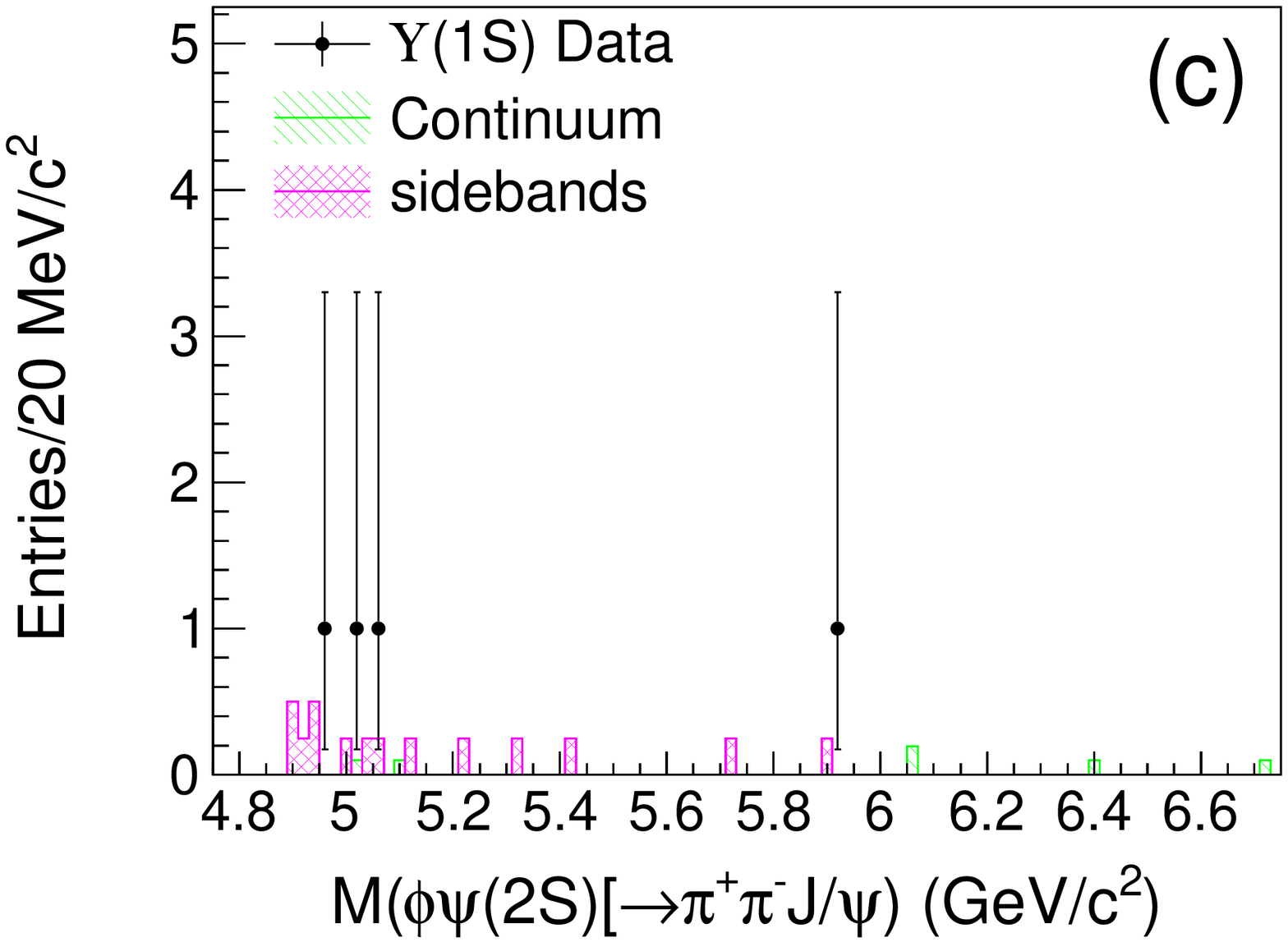}
  \caption{ Invariant mass distributions of the (a) $\phi J/\psi$,
  (b) $\phi \psi(2S)(\to\ell^+\ell^-)$, and (c) $\phi \psi(2S)(\to\pi^+\pi^-J/\psi)$ candidates in $\Upsilon(1S)$
   inclusive decays.
   The points with error bars are events observed at the $\Upsilon(1S)$ peak,
  and the slant-shaded histograms are the scaled continuum contributions from
  the $\sqrt{s}=10.52~\mathrm{GeV}$ continuum data sample
  which overlie the normalized $\psi(2S)$ mass sideband backgrounds
  (the cross-shaded histograms) for the two $\psi(2S)$ decay modes.
  The solid line in panel (a) is the best fit for the $\phi J/\psi$ mass spectrum
  and the dashed line is the total fitted background.}\label{fig-X-PhiJpsi}
\end{figure*}

\begin{figure*}[htbp]
\includegraphics[width=0.34\textwidth]{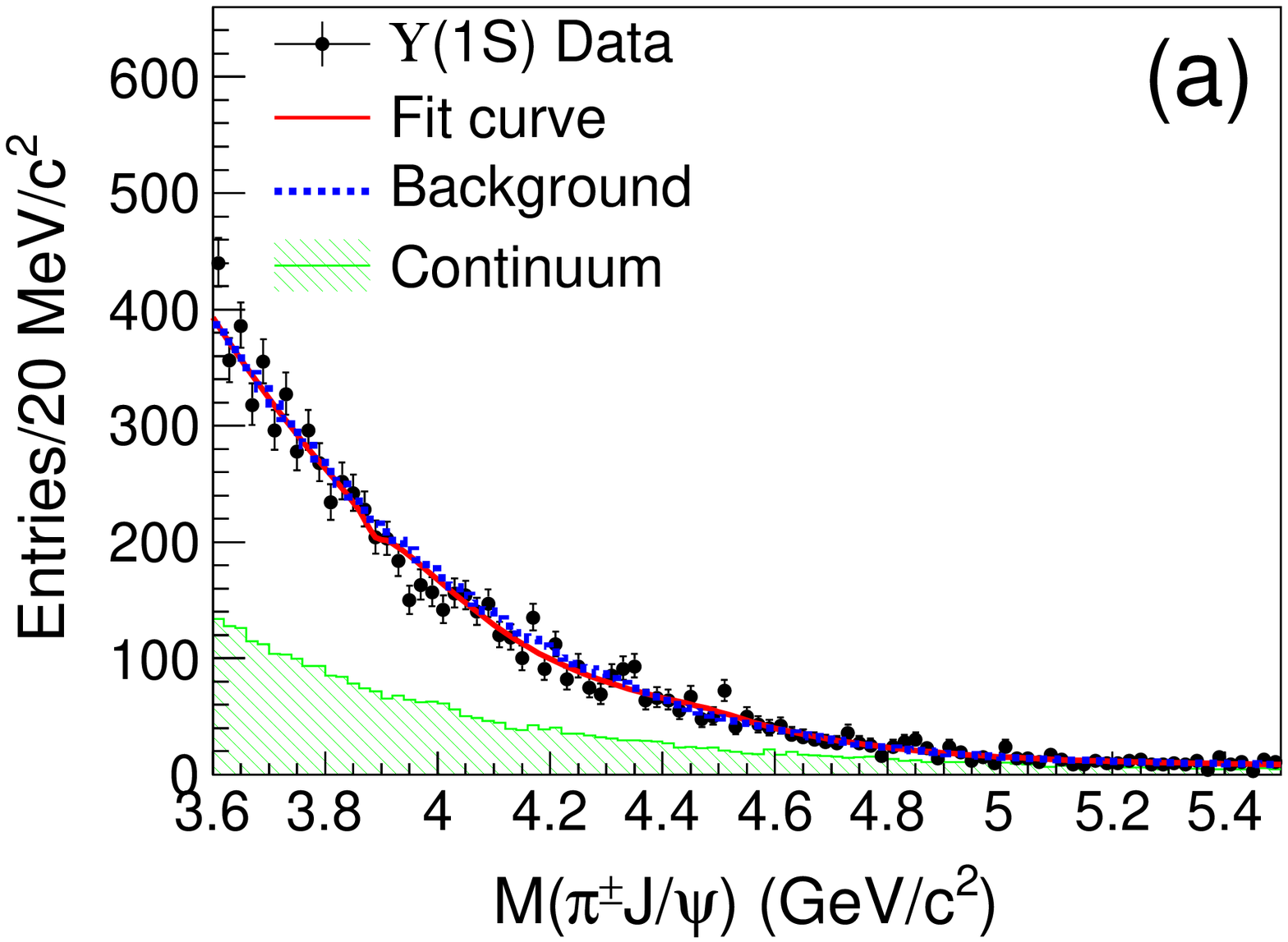} \hspace{-0.45cm}
\includegraphics[width=0.34\textwidth]{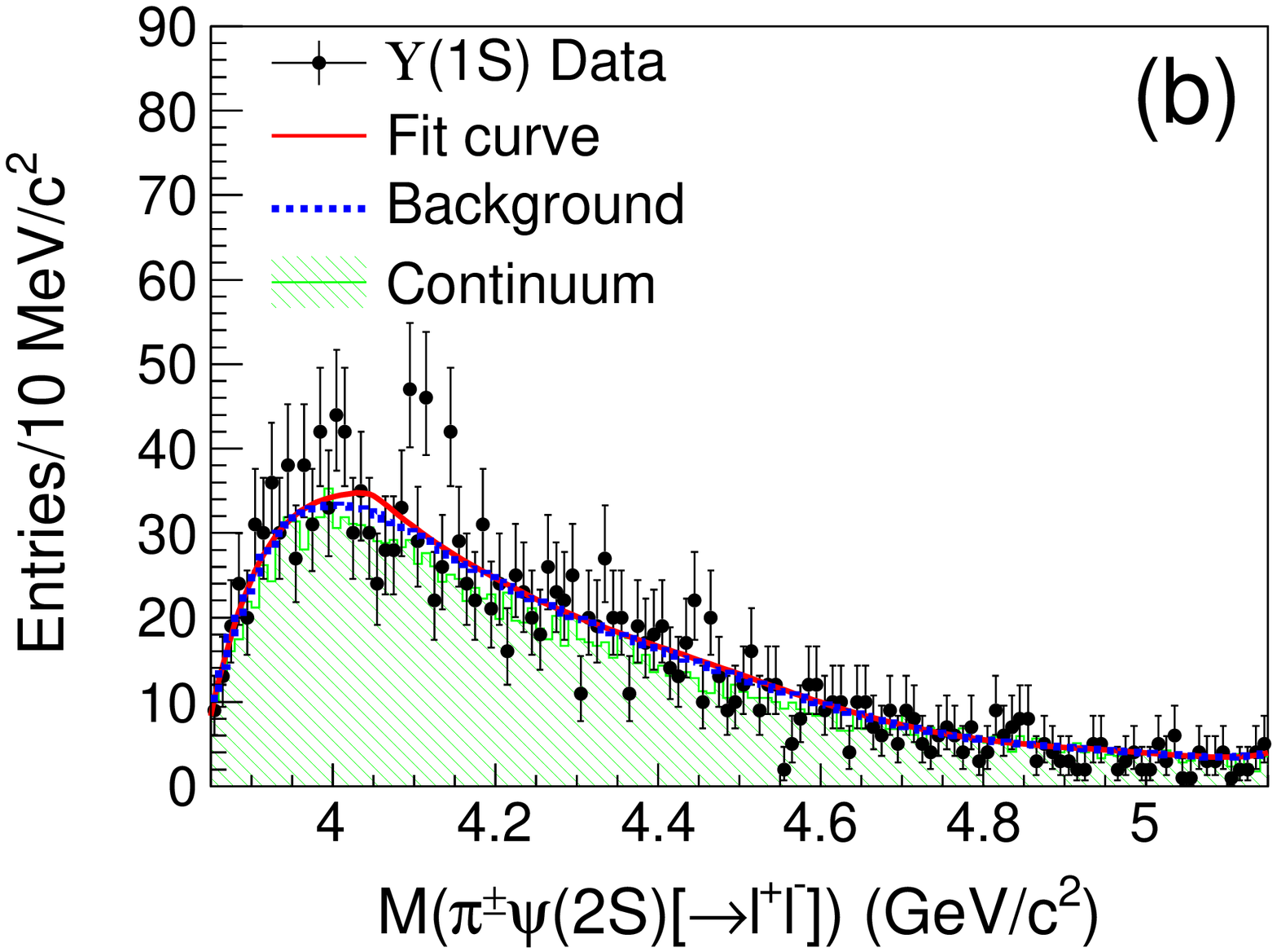} \hspace{-0.45cm}
\includegraphics[width=0.34\textwidth]{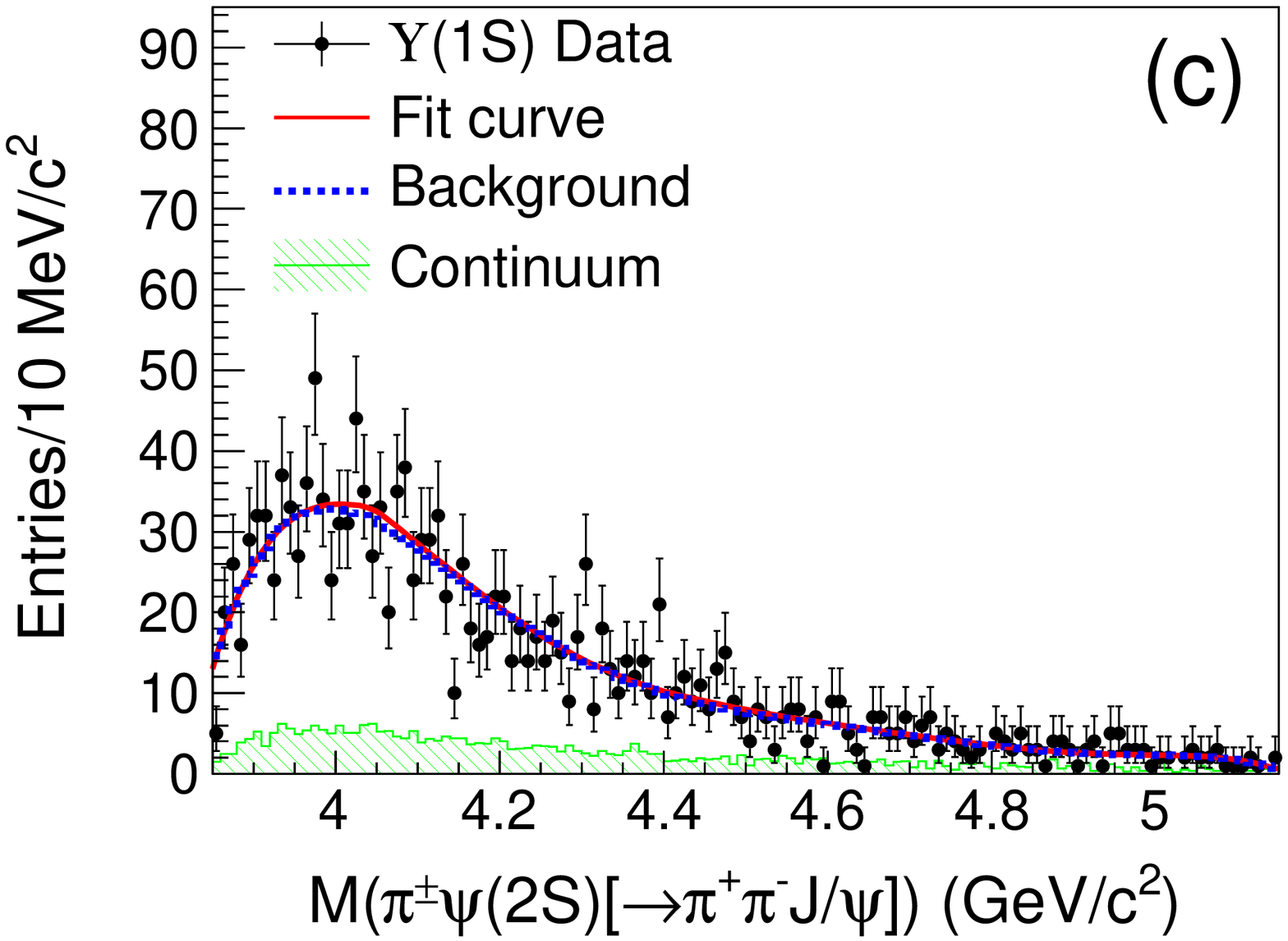}
  \caption{ Invariant mass distributions of the (a) $\pi^\pm J/\psi$,
  (b) $\pi^\pm \psi(2S)(\to\ell^+\ell^-)$, and (c) $\pi^\pm \psi(2S)(\to\pi^+\pi^-J/\psi)$ candidates
  in $\Upsilon(1S)$ inclusive decays.
  The points with error bars are the $\Upsilon(1S)$ events
  and the shaded histograms are the scaled continuum contributions with the data sample collected at $\sqrt{s}=10.52~\mathrm{GeV}$.
  The solid lines are the best fits with the fitted total background components represented by the dashed lines.}\label{fig-X-Y2X3900}
\end{figure*}

\begin{figure*}[htbp]
\includegraphics[width=0.34\textwidth]{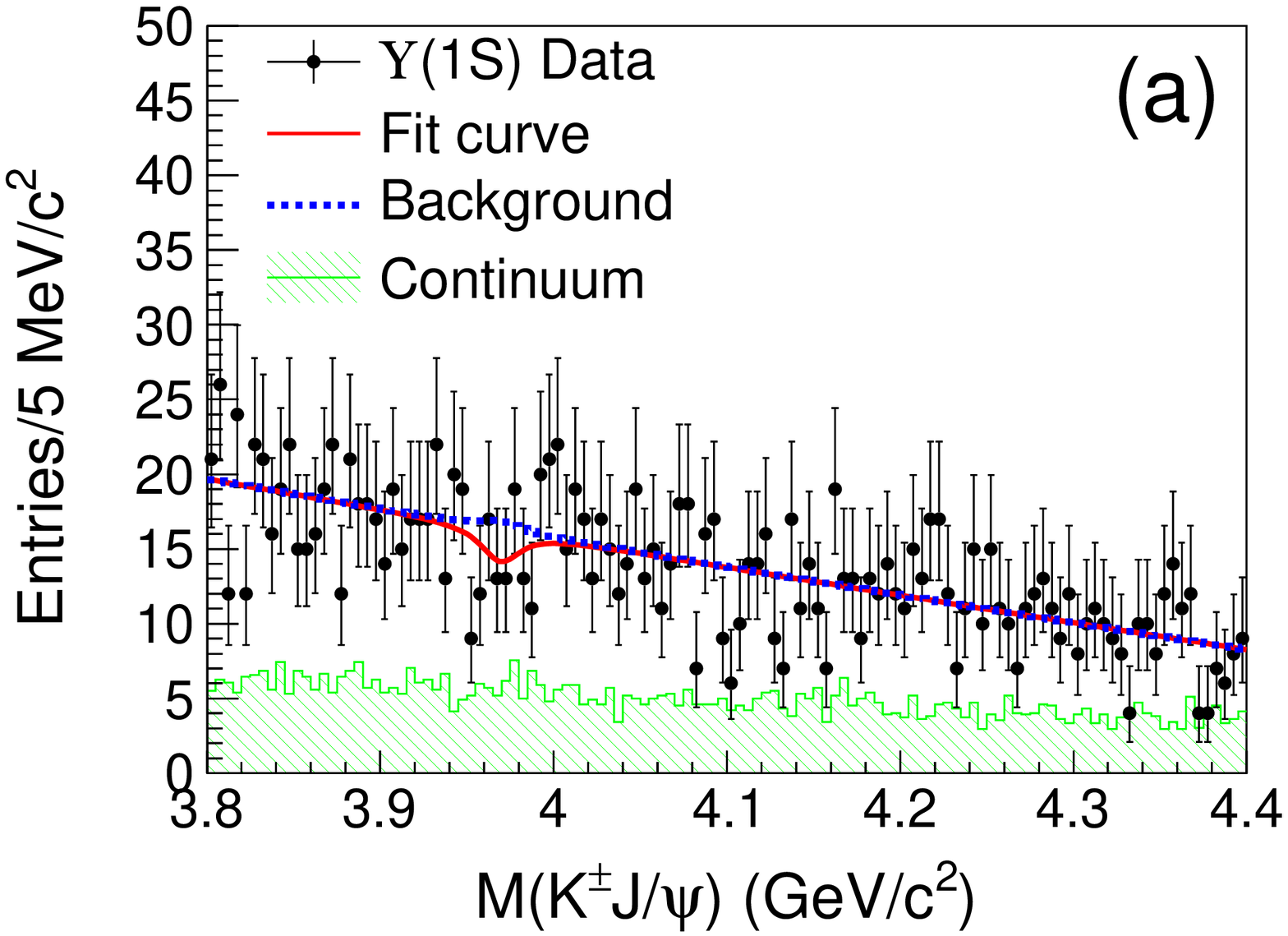} \hspace{-0.45cm}
\includegraphics[width=0.34\textwidth]{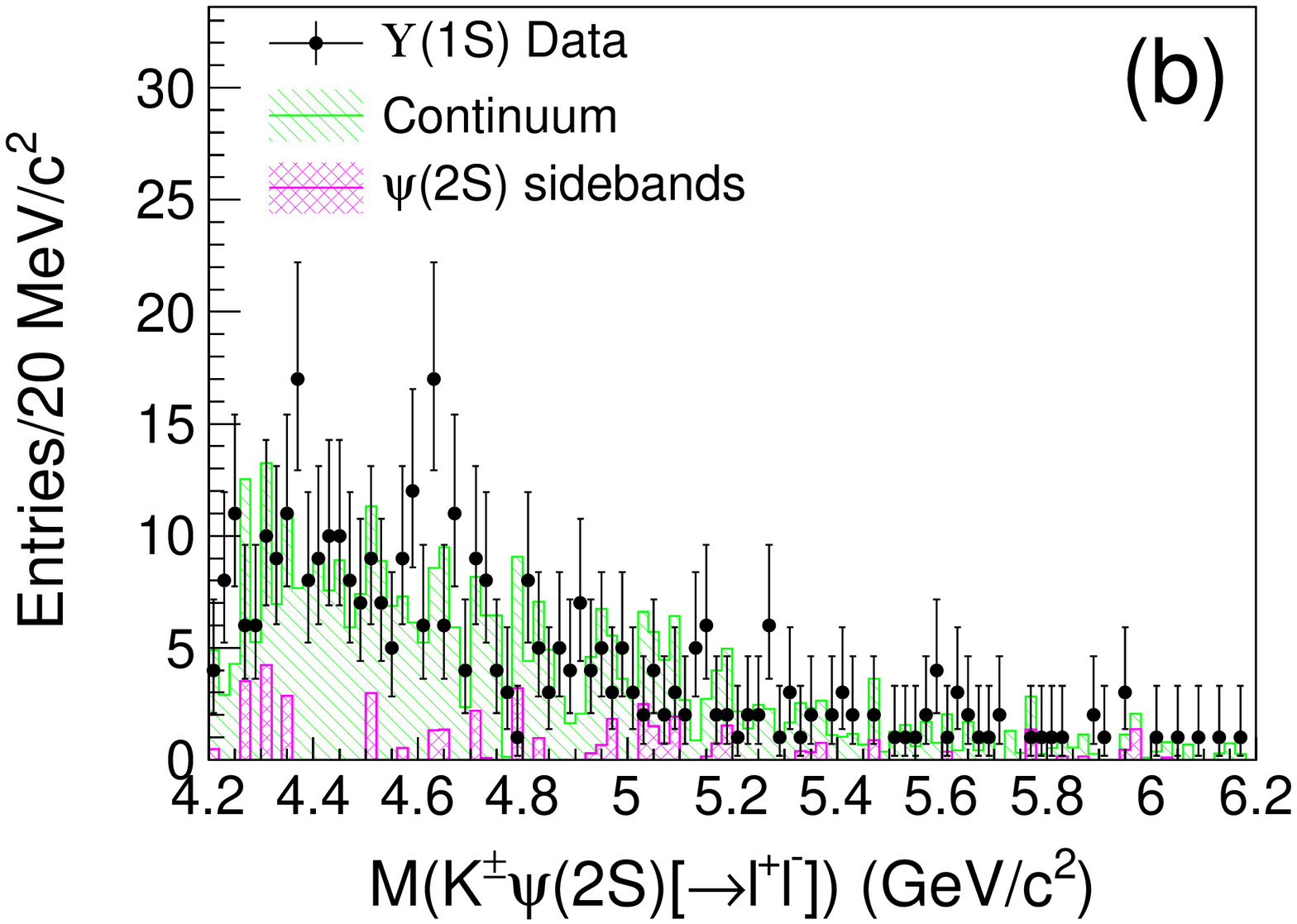} \hspace{-0.45cm}
\includegraphics[width=0.34\textwidth]{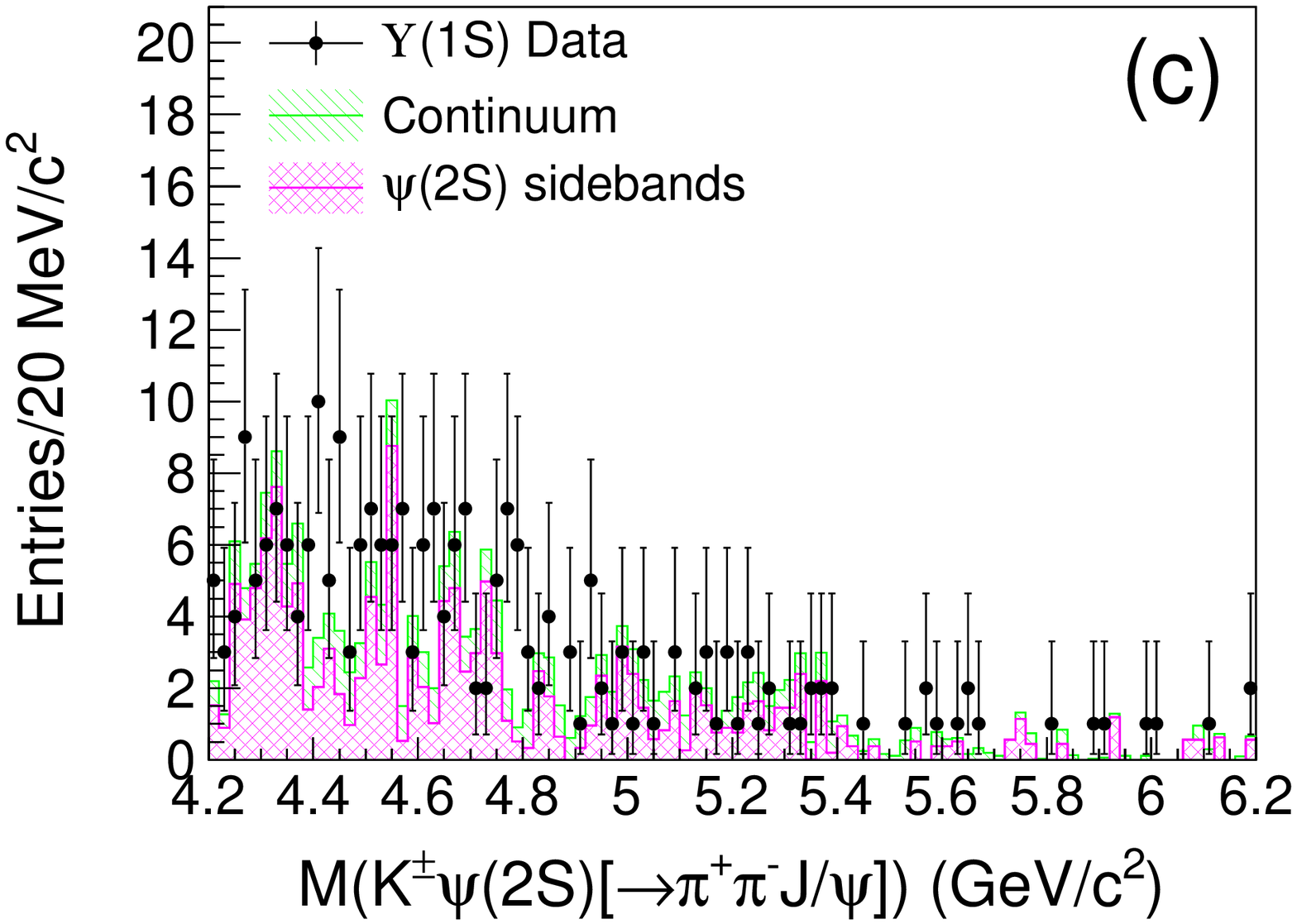}
  \caption{ The (a) $K^\pm J/\psi$,  (b) $K^\pm \psi(2S)(\to\ell^+\ell^-)$, and
   (c) $K^\pm \psi(2S)(\to\pi^+\pi^-J/\psi)$ mass distributions for candidate events
  in the $\Upsilon(1S)$ peak decay sample.
  The points with error bars are the $\Upsilon(1S)$ events
  and the slant-shaded histograms are the scaled continuum contributions
  determined from the data collected at $\sqrt{s}=10.52~\mathrm{GeV}$. The
  normalized $\psi(2S)$ mass-sideband events  are shown as the cross-shaded
  histograms. The solid line in panel (a) is the best fit with the fitted total background component represented
  by the dashed line.
  }\label{fig-X-Y2KJpsi}
\end{figure*}

Figure~\ref{fig-X-Y2X3872} shows the $\pi^+\pi^-J/\psi$ invariant
mass distributions, relevant for the $X(3872)$ and $Y(4260)$ searches,
and those for $\pi^+\pi^-\psi(2S)$, relevant for the $Y(4260)$, $Y(4360)$ and $Y(4660)$.
There are no evident signals for any of these states; the solid lines indicate
the best fit results from a simultaneous fit to the $\Upsilon(1S)$
and continuum data samples. The dashed curves are the total
background estimates.
% described by the sum of the Chebychev
%polynomial functions and normalized continuum backgrounds.
The same representations of the curves and histograms are used for the
$K^+K^-J/\psi$ and $\phi J/\psi$ mass distributions shown in
Figs.~\ref{fig-X-KKJpsi}(a) and ~\ref{fig-X-PhiJpsi}(a), respectively, and for
the charged $\pi^{\pm}J/\psi(\psi(2S))$ and $K^\pm J/\psi$ modes
in Figs.~\ref{fig-X-Y2X3900} and~\ref{fig-X-Y2KJpsi}(a),
respectively.

Because of the large difference between the $X(3872)$ and
$Y(4260)$ widths~\cite{ChinPhysC38.090001}, the fit range for the
$M(\pi^+\pi^-J/\psi)$  spectrum is separated into low and high
mass regions with different bin widths as shown in
Figs.~\ref{fig-X-Y2X3872}(a)~and~(b). The sharp peak at the
$\psi(2S)$ nominal mass, as seen in Fig.~\ref{fig-X-Y2X3872}(a),
is from  $\Upsilon(1S)\to\psi(2S)+{\rm anything} \to\pi^+\pi^-J/\psi+{\rm anything}$.
In contrast, no $X(3872)$ signal is observed.
Using the MC-determined
$\psi(2S)$ signal shape, the fit
yields $139.8\pm20.2$ $\psi(2S)$ signal events.
%with a statistical significance of $8.5\sigma$.
With the MC-determined reconstruction efficiency
($0.98\%$), the resulting branching fraction of
the $\Upsilon(1S)$ inclusive decay into $\psi(2S)$ is
$(1.39\pm0.20({\rm stat.})\pm0.13({\rm syst.}))\times10^{-4}$.
The measurement is in agreement with that listed in
Table~\ref{Table-Nfit-Psi}, where the $\psi(2S)$ candidates are
reconstructed via $\ell^+\ell^-$ and $\pi^+\pi^-J/\psi$. In
addition, there is no evidence for $Y(4260)$ signal in the
$\pi^+\pi^-J/\psi$ mass spectrum shown in
Fig.~\ref{fig-X-Y2X3872}(b). We also search for the $Y(4260)$
state in the $\pi^+\pi^-\psi(2S)$ mass spectra shown in
Figs.~\ref{fig-X-Y2X3872}(c) and \ref{fig-X-Y2X3872}(d) for the
$\ell^+\ell^-$ and $\pi^+\pi^+J/\psi$
decay modes, respectively, of the $\psi(2S)$ candidates, as well as the $Y(4360)$ and $Y(4660)$
states. No enhancements near the nominal masses of these states are
evident.

The $Y(4260)$ has been seen in the $K^+K^-J/\psi$ channel by
CLEO-c~\cite{Coan:2006rv}. Figure~\ref{fig-X-KKJpsi}(a) shows the
$K^+K^-J/\psi$ invariant mass distributions for the candidate
$\Upsilon(1S)$ inclusive decays. The fit to the spectrum of
$M(K^+K^-J/\psi)$ is performed above $4.10~\mathrm{GeV}/c^2$, which
is somewhat above the $K^+K^-J/\psi$ mass threshold of 
$4.085~\mathrm{GeV}/c^2$.
The invariant mass distributions of the
$K^+K^-\psi(2S)$ candidates in $\Upsilon(1S)$ inclusive decays are
shown in Figs.~\ref{fig-X-KKJpsi}(b) and
\ref{fig-X-KKJpsi}(c) for $\psi(2S)\to\ell^+\ell^-$ and
$\pi^+\pi^-J/\psi$, respectively. The slant-shaded histograms (the
scaled continuum backgrounds) overlie the cross-shaded ones that
represent the normalized $\psi(2S)$ mass sideband. No
evidence is found for new structures or any of the known $XYZ$
states. The $Y(4140)$ and $X(4350)$ states have been reported in
the $\phi J/\psi$ decay channel by CDF~\cite{Aaltonen:2009tz} and Belle~\cite{PhysRevLett.104.112004}. Figure~\ref{fig-X-PhiJpsi} shows
the $\phi J/\psi$ and $\phi \psi(2S)$ invariant mass
distributions, where the few events that survive do not appear to
have any statistically significant clustering near 4140~MeV/$c^2$,
4350~MeV/$c^2$ nor any other mass. The results of a fit to
$M(\phi J/\psi)$ in Fig.~\ref{fig-X-PhiJpsi}(a) are shown as a solid curve.
Figures~\ref{fig-X-PhiJpsi}(b) and \ref{fig-X-PhiJpsi}(c) show the
$\phi\psi(2S)$ invariant mass distributions; there are only $7$
and $4$ events that survive in the $\ell^+\ell^-$ and
$\pi^+\pi^-J/\psi$ decay modes, respectively.  No structures are identified.

We search for various charged $Z_c^\pm$ states decaying into $\pi^\pm
J/\psi(\psi(2S))$. Figure~\ref{fig-X-Y2X3900} shows
the $\pi^\pm J/\psi$, $\pi^\pm\psi(2S)(\to\ell^+\ell^-)$,
and $\pi^\pm\psi(2S)(\to\pi^+\pi^-J/\psi)$ invariant mass
distributions for the $\Upsilon(1S)$ peak data as well as the fit
ranges and results.
For all three channels, the background events represent the
$\Upsilon(1S)$ data well, indicating insignificant production of any $Z_c^\pm$ states.
We do not observe any $Z_c^\pm(3900)$, $Z_c^\pm(4200)$ or $Z_c^\pm(4430)$
signals in the $\pi^\pm J/\psi$ mode nor any $Z_c^\pm(4050)$ or $Z_c^\pm(4430)$ signals in the $\pi^\pm \psi(2S)$ mode.
We search for the predicted
$Z_{cs}^\pm(\to K^\pm J/\psi)$ state
---the strange partner of $Z_c^{\pm}(3900)$~\cite{JKPS55.424,PhysRevD.88.096014}---
with mass
$M=(3.97\pm0.08)~\mathrm{GeV}/c^2$ and width
$\Gamma=(24.9\pm12.6)~\mathrm{MeV}$ in $\Upsilon(1S)$ inclusive
decays. The invariant mass distribution of the $K^\pm J/\psi$ candidates is
presented in Fig.~\ref{fig-X-Y2KJpsi}(a). No evidence for such a
structure is seen near the predicted $Z_{cs}^\pm$  mass.
The signal significance from the fit is less than
2$\sigma$. A fit with a Breit-Wigner that
interferes with a smooth background function
yields a signal significance of only 1.2$\sigma$. In the
$K^\pm\psi(2S)$ mode, no exotic $XYZ$ states have been seen nor predicted. For completeness, we present
the invariant mass distributions of the $K^\pm\psi(2S)$ candidates with the
$\psi(2S)$ decays into the $\ell^+\ell^-$ and $\pi^+\pi^-J/\psi$
final states  in Figs.~\ref{fig-X-Y2KJpsi}(b) and
\ref{fig-X-Y2KJpsi}(c), respectively. The sum of the normalized
continuum and sideband backgrounds agrees well with the
data.

The fitted signal yields ($N_{\rm fit}$) of the $XYZ$ states that
are considered in this analysis are presented  in
Table~\ref{Table-X-Summary}. Since  the statistical
significance in each case is less than $3\sigma$, upper limits on the number of
signal events, $N_{\rm up}$, are determined at the $90\%$
credibility level (C.L.) by solving the equation $\int^{N_{\rm
up}}_0\mathcal{L}(x)dx/\int^{+\infty}_0\mathcal{L}(x)dx=0.9$~\cite{cl},
where $x$ is the number of fitted signal events and
$\mathcal{L}(x)$ is the likelihood function in the fit to data.
To take into account systematic uncertainties (discussed below),
the above likelihood is convolved with a Gaussian function whose
width equals the total systematic uncertainty. The calculated
upper limits on the number of signal events ($N_{\rm up}$) and the
branching fraction ($\mathcal{B}$) for each state are listed in
Table~\ref{Table-X-Summary}, together with the reconstruction
efficiencies ($\varepsilon$), the systematic uncertainties
($\sigma_{\rm syst}$), and the signal significances ($\Sigma$);
the latter are calculated using
$\sqrt{-2\ln(\mathcal{L}_0/\mathcal{L}_{\rm max})}$, where
$\mathcal{L}_0$ and $\mathcal{L}_{\rm max}$ are the likelihoods of
the fits without and with a signal component, respectively.

\begin{table*}[htbp]
  \caption{\label{Table-X-Summary} Summary of the upper limits on the
  $\Upsilon(1S)$ inclusive decays into the exotic charmoniumlike states $XYZ$,
  where $N_{\rm fit}$ is the number of fitted signal events,
  $N_{\rm up}$ is the upper limit on the number of signal events taking into account systematic errors,
  $\varepsilon$ is the reconstruction efficiency,
  $\sigma_{\rm syst}$ is the total systematic uncertainty,
  $\Sigma$ is the signal significance with systematic errors included,
  and $\mathcal{B}_{R}^{\rm prod}=\mathcal{B}(\Upsilon(1S)\to XYZ+ {\rm anything})\mathcal{B}(XYZ\to J/\psi(\psi(2S))+ {\rm hadrons})$  is the
  measured product branching fraction at the 90\% C.L.}
  \begin{tabular}{lr@{$\pm$}lr@{.}lcccc}
  \hline\hline
    State &\multicolumn{2}{c}{$N_{\rm fit}$} &\multicolumn{2}{c}{$N_{\rm up}$} &$\varepsilon(\%)$ &$\sigma_{\rm syst}(\%)$ &$\Sigma(\sigma)$ &$\mathcal{B}_{R}^{\rm prod}$ \\
  \hline
    $X(3872)\to\pi^+\pi^-J/\psi$   & $4.8$   &$15.4$ & 31 & 4 & $3.26$ & $18.7$ & $0.3$ & $<9.5\times10^{-6}$ \\
    $Y(4260)\to\pi^+\pi^-J/\psi$   & $-31.1$ &$88.9$ & $134$&$6$ & $3.50$ & $35.6$ & $-$   & $<3.8\times10^{-5}$ \\
    $Y(4260)\to\pi^+\pi^-\psi(2S)$ & $6.7$   &$29.4$ & $56$ &$9$ & $0.71$ & $35.0$ & $0.2$ & $<7.9\times10^{-5}$ \\
    $Y(4360)\to\pi^+\pi^-\psi(2S)$ & $-25.4$ &$30.1$ & $45$ &$6$ & $0.86$ & $50.0$ & $-$   & $<5.2\times10^{-5}$ \\
    $Y(4660)\to\pi^+\pi^-\psi(2S)$ & $-55.0$ &$26.2$ & $23$ &$1$ & $1.06$ & $40.7$ & $-$   & $<2.2\times10^{-5}$ \\
    $Y(4260)\to K^+K^-J/\psi$      & $-13.7$ &$10.9$ & $14$ &$5$ & $1.91$ & $45.8$ & $-$   & $<7.5\times10^{-6}$ \\
    $Y(4140)\to \phi J/\psi$       & $-0.1$  &$1.2$  & $3$  &$6$ & $0.69$ & $11.0$ & $-$   & $<5.2\times10^{-6}$ \\
    $X(4350)\to \phi J/\psi$       & $2.3$   &$2.5$  & $7$  &$6$ & $0.92$ & $10.4$ & $1.2$ & $<8.1\times10^{-6}$ \\
    $Z_c(3900)^\pm\to\pi^\pm J/\psi$   & $-26.5$ & $39.1$  & $57$ &$5$ & $4.39$ & $47.3$ & $-$   & $<1.3\times10^{-5}$ \\
    $Z_c(4200)^\pm\to\pi^\pm J/\psi$   & $-238.6$& $154.2$ & $235$&$1$ & $3.87$ & $48.4$ & $-$   & $<6.0\times10^{-5}$ \\
    $Z_c(4430)^\pm\to\pi^\pm J/\psi$   & $94.2$  & $71.4$  & $195$&$8$ & $3.97$ & $34.4$ & $1.2$ & $<4.9\times10^{-5}$ \\
    $Z_c(4050)^\pm\to\pi^\pm \psi(2S)$ & $37.0$  & $47.7$  & $112$&$7$ & $1.27$ & $46.2$ & $0.4$ & $<8.8\times10^{-5}$ \\
    $Z_c(4430)^\pm\to\pi^\pm \psi(2S)$ & $23.2$  & $42.4$  & $92$ &$0$ & $1.35$ & $47.1$ & $0.1$ & $<6.7\times10^{-5}$ \\
    $Z_{cs}^\pm\to K^\pm J/\psi$       & $-22.2$ & $17.4$  & $22$ &$4$ & $3.88$ & $48.7$ & $-$   & $<5.7\times10^{-6}$ \\
  \hline\hline
  \end{tabular}
\end{table*}

%%%%%%%%%%%%%%%%%%%%%%%%%%%%%%%%%%%%%%%%%%%%%%%%%%%%%%%%%%%%%%%%%%%%%%%%%%%%%%%%%%%%%%%%%%%%%%%%%%%
%%%%%%%%%%%%%%%%%%%%%%%%%%%%%%%%% Systematic Uncertainties %%%%%%%%%%%%%%%%%%%%%%%%%%%%%%%%

Several sources of systematic errors are taken into account in the
branching fraction measurements. Tracking efficiency uncertainty
is estimated to be $0.35\%$ per track with high momentum and is
additive. Based on the measurements of the identification
efficiencies of lepton pairs from $\gamma\gamma\rightarrow
\ell^+\ell^-$ events and pions from a low-background sample of
$D^\ast$ events,  MC simulation yields uncertainties of
$1.6\%$ for each lepton, $1.4\%$ for each pion, and $1.3\%$ for
each kaon. The trigger efficiency evaluated from
simulation is greater than $99.9\%$ with an uncertainty that is
negligibly small. The difference in the signal yields when the mass
and width of each $XYZ$ state are varied by $1\sigma$ is used as
an estimate of the systematic error associated with mass and
width uncertainties~\cite{ChinPhysC38.090001}. In the simulation
of generic $J/\psi(\psi(2S))$ decays, the unknown decay channels
are produced by the Lund fragmentation model in
PYTHIA~\cite{JHEP2006.026}. By generating different sets of MC
samples with different relative probabilities to produce the
various possible $q\bar{q}$ ($q=u,~d,~s$) pairs in the
$J/\psi(\psi(2S))$ decays, the largest difference in the
efficiencies is found to be less than $0.1\%$ and is neglected.
The errors on the branching fractions of the intermediate states
are taken from the Particle Data Group
tables~\cite{ChinPhysC38.090001}; these are  $1.1\%$, $6.3\%$,
$1.2\%$, and $1.0\%$ for $J/\psi\to \ell^+\ell^-$, $\psi(2S)\to
\ell^+\ell^-$, $\psi(2S)\to\pi^+\pi^-J/\psi$, and $\phi\to
K^+K^-$, respectively; the weighted average for the two
$\psi(2S)$ decay modes is  $3.5\%$. By varying the background
shapes, the order of the Chebyshev polynomial and the fitting
range, the deviations of the fitted signal yields for
$J/\psi(\psi(2S))$ productions are estimated for each $x$ bin.
The upper limits on the signal yields
vary by less than 49.4\%, depending
on the decay mode. The MC statistical errors are estimated using
the reconstruction efficiencies and the number of generated
events;  these are $1.0\%$ or less. The error on the total number
of $\Upsilon(1S)$ events is $2.0\%$. Assuming that all sources
are independent, their uncertainties are summed in quadrature. The total
systematic errors ($\sigma_{\rm syst}$) for each channel are
listed in Table~\ref{Table-X-Summary}.

%%%%%%%%%%%%%%%%%%%%%%%%%%%%%%%%%%%%%%%%%%%%%%%%%%%%%%%%%%%%%%%%%%%%%%%%%%%%%%%%%%%%%%%%%%%%%%%%%%%
%%%%%%%%%%%%%%%%%%%%%%%%%%%%%%%%%%%%%%%%%%% Summary %%%%%%%%%%%%%%%%%%%%%%%%%%%%%%%%%%%%%%%%%%%%%%%
%%%%%%%%%%%%%%%%%%%%%%%%%%%%%%%%%%%%%%%%%%%%%%%%%%%%%%%%%%%%%%%%%%%%%%%%%%%%%%%%%%%%%%%%%%%%%%%%%%%

In summary, using the $102\times10^6$ $\Upsilon(1S)$ events
collected with the Belle detector, distinct $J/\psi$ and
$\psi(2S)$ signals are observed in the $\Upsilon(1S)$ inclusive
decays. The corresponding branching fractions are measured to be
$\mathcal{B}(\Upsilon(1S)\to J/\psi+ {\rm anything})=(5.25\pm0.13(\mathrm{stat.})\pm0.25(\mathrm{syst.}))\times10^{-4}$
and
$\mathcal{B}(\Upsilon(1S)\to\psi(2S)+ {\rm anything})=(1.23\pm0.17(\mathrm{stat.})\pm0.11(\mathrm{syst.}))\times10^{-4}$
with substantially improved precision compared to previous results of
$(6.5\pm 0.7)\times 10^{-4}$~\cite{PhysRevD.70.072001, PhysLettB224.445}
and $(2.7\pm 0.9)\times
10^{-4}$~\cite{PhysRevD.70.072001} for $J/\psi$ and $\psi(2S)$, respectively.
%over previous work.
Several theoretical papers have suggested the study of $J/\psi$ production
in $\Upsilon(1S)$ decays as an example of charmonium production
mechanisms in gluon-rich environments.  Some
color-octet~\cite{octet} and color-singlet~\cite{signlet} models
predict $\BR(\Upsilon(1S)\to J/\psi+{\rm anything}$) of $6.2\times 10^{-4}$ and
$5.9\times 10^{-4}$, respectively.
Our measured value is of the same order as the theoretical estimations.
We also search for a variety
of $XYZ$ states in $\Upsilon(1S)$ inclusive decays for the first
time, where the $XYZ$ candidates of interest are reconstructed
from their final states that contain a $J/\psi(\psi(2S))$ and up
to two charged light hadrons ($K^\pm/\pi^\pm$). No evident signal
is found for any of them and $90\%$ C.L. upper limits are set on
the product branching fractions and listed in
Table~\ref{Table-X-Summary}. There is no striking evidence for
previously unseen structures in $K^+K^-\psi(2S)$ and
$K^\pm\psi(2S)$ invariant mass distributions.

%%%%%%%%%%%%%%%%%%%%%%%%%%%%%%%%%%%%%%%%%%%%%%%%%%%%%%%%%%%%%%%%%%%%%%%%%%%%%%%%%%%%%%%%%%%%%%%%%%%
%%%%%%%%%%%%%%%%%%%%%%%%%%%%%%%%%%%%% Acknowledgement %%%%%%%%%%%%%%%%%%%%%%%%%%%%%%%%%%%%%%%%%%%%%
%%%%%%%%%%%%%%%%%%%%%%%%%%%%%%%%%%%%%%%%%%%%%%%%%%%%%%%%%%%%%%%%%%%%%%%%%%%%%%%%%%%%%%%%%%%%%%%%%%%
%----------- Long version, for most papers -----------

We thank the KEKB group for the excellent operation of the
accelerator; the KEK cryogenics group for the efficient
operation of the solenoid; and the KEK computer group,
the National Institute of Informatics, and the
PNNL/EMSL computing group for valuable computing
and SINET4 network support.  We acknowledge support from
the Ministry of Education, Culture, Sports, Science, and
Technology (MEXT) of Japan, the Japan Society for the
Promotion of Science (JSPS), and the Tau-Lepton Physics
Research Center of Nagoya University;
the Australian Research Council;
Austrian Science Fund under Grant No.~P 22742-N16 and P 26794-N20;
the National Natural Science Foundation of China under Contracts
No.~10575109, No.~10775142, No.~10875115, No.~11175187, No.~11475187
and No.~11575017;
the Chinese Academy of Science Center for Excellence in Particle Physics;
the Ministry of Education, Youth and Sports of the Czech
Republic under Contract No.~LG14034;
the Carl Zeiss Foundation, the Deutsche Forschungsgemeinschaft, the
Excellence Cluster Universe, and the VolkswagenStiftung;
the Department of Science and Technology of India;
the Istituto Nazionale di Fisica Nucleare of Italy;
the WCU program of the Ministry of Education, National Research Foundation (NRF)
of Korea Grants No.~2011-0029457,  No.~2012-0008143,
No.~2012R1A1A2008330, No.~2013R1A1A3007772, No.~2014R1A2A2A01005286,
No.~2014R1A2A2A01002734, No.~2015R1A2A2A01003280 , No. 2015H1A2A1033649;
the Basic Research Lab program under NRF Grant No.~KRF-2011-0020333,
Center for Korean J-PARC Users, No.~NRF-2013K1A3A7A06056592;
the Brain Korea 21-Plus program and Radiation Science Research Institute;
the Polish Ministry of Science and Higher Education and
the National Science Center;
the Ministry of Education and Science of the Russian Federation and
the Russian Foundation for Basic Research;
the Slovenian Research Agency;
Ikerbasque, Basque Foundation for Science and
the Euskal Herriko Unibertsitatea (UPV/EHU) under program UFI 11/55 (Spain);
the Swiss National Science Foundation;
the Ministry of Education and the Ministry of Science and Technology of Taiwan;
and the U.S.\ Department of Energy and the National Science Foundation.
This work is supported by a Grant-in-Aid from MEXT for
Science Research in a Priority Area (``New Development of
Flavor Physics'') and from JSPS for Creative Scientific
Research (``Evolution of Tau-lepton Physics'').

%%%%%%%%%%%%%%%%%%%%%%%%%%%%%%%%%%%%%
%%%%%%%%%%%%%%%%%%%%%%%%%%%%%%%%%%%%%
%\bibliography{Ref_Y2XYZ}

\begin{thebibliography}{**}
\bibitem{cl} In common high energy physics usage, this
Bayesian interval has been reported as ``confidence interval,'' which
is a frequentist-statistics term.
\end{thebibliography}

\end{document}